\documentclass[onecolumn,apj]{emulateapj}
\newcommand\bl{{\mathbf l}} 
\newcommand\bm{{\mathbf m}}

\newcommand\bs{{\mathbf s}}

\newcommand\bB{{\mathbf B}}

\newcommand\bP{{\mathbf P}}
\newcommand\bQ{{\mathbf Q}}

\newcommand\bmu{\boldsymbol{\mu}}

\newcommand\norm[1]{\lVert #1 \rVert}

\usepackage{amsmath,amsfonts}
\usepackage{lscape}
\shorttitle{High precision orbital and physical parameters of SB2s}
\shortauthors{M. Konacki, M. Muterspaugh, S. Kulkarni and K. Helminiak}
\begin{document}
\title{High precision orbital and physical parameters of double-lined
spectroscopic binary stars - HD78418, HD123999, HD160922, HD200077 
and HD210027}
\author{Maciej Konacki\altaffilmark{1,2,3}, Matthew W.
Muterspaugh\altaffilmark{4,5}, Shrinivas R. Kulkarni\altaffilmark{6}
and Krzysztof G. He{\l}miniak\altaffilmark{1}}

\altaffiltext{1}{Nicolaus Copernicus Astronomical Center, Polish Academy of
Sciences, Rabianska 8, 87-100 Torun, Poland}
\altaffiltext{2}{Astronomical Observatory, A. Mickiewicz University,
Sloneczna 36, 60-286 Poznan, Poland}
\altaffiltext{3}{e-mail: maciej@ncac.torun.pl}
\altaffiltext{4}{Department of Mathematics and Physics, College of Arts
and Sciences, Tennessee State University, Boswell Science Hall,
Nashville, TN 37209, USA}
\altaffiltext{5}{Tennessee State University, Center of Excellence in Information
Systems, 3500 John A. Merritt Blvd., Box No. 9501,
Nashville, TN 37203-3401, USA}
\altaffiltext{6}{Division of Physics, Mathematics and Astronomy, California Institute
of Technology, Pasadena, CA 91125, USA}


\begin{abstract}
We present high precision radial velocities (RVs) of double-lined spectroscopic
binary stars HD78418, HD123999, HD160922, HD200077 and HD210027. They were
obtained based on the high resolution echelle spectra collected with the Keck I/Hires, 
Shane/CAT/Hamspec and TNG/Sarge telescopes/spectrographs over the years
2003-2008 as a part of TATOOINE search for circumbinary planets. 
The RVs were computed using our novel iodine cell technique for double-line
binary stars which relies on tomographically disentangled spectra of the
components of the binaries. The precision of the RVs is of the order of
1-10 m~s$^{-1}$ and to properly model such measurements one
needs to account for the light time effect within the binary's orbit,
the relativistic effects and the RV variations due to the tidal 
distortions of the components of the binaries.
With such proper modeling, our RVs combined with the archival visibility 
measurements from the Palomar Testbed Interferometer allow us to derive very
precise spectroscopic/astrometric orbital and physical parameters of the 
binaries. In particular, we derive the masses, the absolute K and H band 
magnitudes and the parallaxes. The masses together with the absolute 
magnitudes in the K and H bands enable us to estimate the ages of the 
binaries.

These RVs allow us to obtain some of the most accurate mass determinations of 
binary stars. The fractional accuracy in $m\sin i$ only and hence based on 
the RVs alone ranges from 0.02\% to 0.42\%. When combined with the PTI astrometry, 
the fractional accuracy in the masses ranges in the three best cases from 0.06\% to 
0.5\%.  Among them, the masses of HD210027 components rival in precision the mass 
determination of the components of the relativistic double pulsar system PSR~J0737-3039. 
In the near future, for double-lined eclipsing binary stars we expect to 
derive masses with a fractional accuracy of the order of up to $\sim$0.001\% with our 
technique. This level of precision is an order of magnitude higher than 
of the most accurate mass determination for a body outside the Solar System
--- the double neutron star system PSR~B1913+16.
\end{abstract}


\section{Introduction}

The first observations of a spectroscopic binary star, $\zeta$ UMa (Mizar), 
were announced by Edward C. Pickering (1846-1919) on 13 November 1889 during 
a meeting of the National Academy of Sciences in Philadelphia
\citep{Pickering:90::}. A similar 
announcement about $\beta$ Per (Algol) was made by Herman C. Vogel (1841-1907) 
on 28 November 1889 during a session of Konglich-Preussiche Akademie der 
Wissenschaften \citep{Vogel:90a::,Vogel:90b::}. 
Even though a transatlantic telegraph cable had been available since the 1860s, it was
quite unusual timing for a pre-astro-ph era. Pickering 
noted that the K line of Mizar occasionally appeared double and this way 
discovered the first double-lined spectroscopic binary star. Vogel measured
radial velocities (RVs) of Algol and used them to prove that the known 
variations in the brightness of Algol are indeed caused by 
a ``dark satellite revolving about it" \citep{Vogel:90b::}.

Vogel and collaborators built a series of prism spectrographs for 
the 30-cm refractor of the Potsdam Astrophysical Observatory
\citep{Vogel:00::}. In 1888, they initiated 
a photographic radial-velocity program. Vogel was not the first to 
take a photograph of a stellar spectrum but improved the 
technique and obtained an RV precision of 2-4 km~s$^{-1}$ \citep{Vogel:91::}.
Vogel (1906) and Pickering (1908) were both awarded the Bruce medal. 
Yet another Bruce medalist (1915), William W. Campbell (1862-1938) 
and collaborators carried out a large spectroscopic program at the Lick Observatory
and discovered many spectroscopic binaries \citep{Camp:11::}. They prepared the
first catalog of spectroscopic binaries \citep{Camp:05::}. Vogel was the director
of the Potsdam Observatory for 25 years, Pickering of the Harvard 
College Observatory for 42 years and Campbell of the Lick 
Observatory for 30 years. The administrative duties
must have been not too distracting those days.

Both Vogel and Campbell recognized the importance of flexure and 
temperature control of a spectrograph \citep{Vogel:90c::,Camp:98::,Camp:00::} 
and Campbell also noted the issue of a slit illumination and 
its impact on RV precision \citep{Camp:16::}. Today 
a high RV precision is achieved by either using a very stable fiber 
fed spectrograph that is contained in a controlled environment (a vacuum
tank) or using an absorption cell to superimpose a reference spectrum onto 
a stellar spectrum and this way measure and account for the systematic RV 
errors. Current state of the art precision is at the level of 
$\sim$1 m~s$^{-1}$. It is however important to note that such a precision 
refers to single stars or at best single-lined spectroscopic binaries
where the influence of the secondary spectrum can be neglected.
In such a case, given a stable spectrograph, an RV measurement is
essentially a measurement of a shift of an otherwise constant
shape (spectrum).

Radial velocities (RVs) of double-lined spectroscopic binary stars (SB2) can be 
used effectively to derive basic parameters of stars if the stars happen to be 
eclipsing or their astrometric relative 
orbit can be determined. It is quite surprising that the RV precision of double-lined
binary stars on the average has not improved much over the
last 100 years (see Fig.~\ref{fig1}). With the exception of our previous 
work \citep{Konacki:05::,Konacki:09a::}, the RV precision for such targets 
typically varies from $\sim$0.1  km~s$^{-1}$ to $\sim$1 km~s$^{-1}$ and clearly 
is much worse than what has been achieved for stars with planets or 
single-lined binary stars. The main problem with double-lined binary stars is that 
one has to deal with two sets of superimposed spectral lines whose corresponding 
radial velocities change considerably with typical amplitudes 
of $\sim$ 50-100 km~s$^{-1}$. In consequence a spectrum is
highly variable and obviously one cannot measure RVs by noting 
a simple shift. 

We have developed a novel iodine cell based
approach that employs a tomographic disentangling of the component
spectra of SB2s and allows one to measure RVs of the components
of SB2s with a precision of the order 1-10 m$\,$s$^{-1}$ 
\citep{Konacki:09a::,Konacki:09b::}. Such quality RVs not only enable us
to search for circumbinary extrasolar planets \citep{Konacki:09b::}
but also to determine basic parameters of stars with an unprecedented
precision. In particular the masses of stars for non eclipsing SB2s can 
be easily determined with a fractional accuracy of the order of at least $\sim$$0.1\%$ 
and often even $\sim$$0.01\%$. Moreover, we expect that the accuracy
in masses will reach $\sim$$0.001\%$ level when our method
is applied to eclipsing binary stars. Such a level of precision
is an order of magnitude higher than of the most
accurate mass determination for a body outside the Solar System
--- the double neutron star system PSR~B1913+16 \citep{Nice:08::}.

Below we present our precision RV data sets for five targets
HD78418, HD123999, HD160922, HD200077 and HD210027 from our on-going 
TATOOINE (The Attempt To Observe Outer-planets In Non-single-stellar 
Environments) RV program to search for circumbinary planets. 
All of them have been extensively
observed with the Palomar Testbed Interferometer \citep[PTI;]{Col:99a::}. 
The archival PTI visibility measurements can be used to derive
relative astrometric orbits of the binaries. These combined
with our spectroscopic orbits allow for a complete orbital
and physical description of the systems (with the exception
of the radii of the components). In \S{2} we describe the RV
measurements and their modeling and in \S{3} the visibility
measurements and their modeling. In \S{4} we present
the spectroscopic and astrometric orbital solutions and the resulting
orbital and physical parameters of the binaries. A discussion
is provided in \S{5}.

\section{Radial velocities}

\subsection{Iodine absorption cell and spectroscopic binary stars}

In the iodine cell (I$_{2}$) technique, the Doppler shift of a star
spectrum $\Delta\lambda$ is determined by solving the following
equation \citep{Mar:92::}.
\begin{equation}
\label{i2::}
I(\lambda) =
[F(\lambda+\Delta\lambda_{s})\,T(\lambda+\Delta\lambda_{I_{2}})]
\,\otimes\,PSF
\end{equation}
where $\Delta\lambda_{s}$ is the shift of the star spectrum,
$\Delta\lambda_{I_{2}}$ is the shift of the iodine transmission
function $T$, $\otimes$ represents a convolution and   
$PSF$ a spectrograph's point spread function (PSF). The parameters 
$\Delta\lambda_{s}, \Delta\lambda_{I_{2}}$ as well as parameters 
describing the PSF are determined by performing a least-squares 
fit to the observed (through the iodine cell) spectrum $I$. 
For this purpose, one also needs (1) a high SNR star spectrum taken 
without the cell, $F$ (the intrinsic stellar spectrum), which serves 
as a template for all the spectra observed through the cell and (2) 
the I$_2$ transmission function $T$ obtained with a Fourier
Transform Spectrometer such as the one at the Kitt Peak National Observatory.
The Doppler shift of a star spectrum is then given by
$\Delta\lambda = \Delta\lambda_{s} - \Delta\lambda_{I_{2}}$.
Such an iodine technique can only be applied to single stars. 
This is dictated by the need to supply an observed template spectrum 
of a star in Eq. \ref{i2::}. In the 
case of SB2s, it cannot be accomplished since their spectra are 
always composite and time variable.

We can measure precise radial velocities of both components of an
SB2 with an I$_2$ absorption cell by performing the following steps. 
First, contrary to the standard approach for single stars, we always 
take two subsequent exposures of a binary target --- one with and 
the other without the I$_2$ cell. This way we 
obtain an instantaneous template which is used to model only the adjacent 
exposure taken with the cell. Next, we perform the usual least-squares 
fit and obtain the parameters described in Eq.~\ref{i2::}. Obviously, 
the derived Doppler shift, $\Delta\lambda_i$ (where $i$ denotes the 
epoch of the observation), carries no meaning since each time a different 
template is used (besides it describes a Doppler ``shift" of a composite 
spectrum different at each epoch). However, the parameters 
(in particular the wavelength solution and the parameters describing PSF) 
are accurately determined and can be used to extract the star spectrum, 
$I^{\star,t}(\lambda)$, for each epoch $t$, by inverting 
the equation~\ref{i2::}
\begin{equation}
\label{met::}
I^{\star,t}(\lambda) = [I^{t}(\lambda)\,\otimes^{-1}\,PSF^{t}]
/T(\lambda)
\end{equation}
where $\otimes^{-1}$ denotes deconvolution and $PSF^{t}$, symbolically, the 
set of parameters describing PSF at the epoch $t$. Such a star spectrum has
an accurate wavelength solution, is free of the I$_2$ lines and the influence
of a varying PSF. In the final step, the velocities of both components
of a binary target can be measured with the well known two-dimensional
cross-correlation technique TODCOR \citep{Zuc:94::} 
using as templates the synthetic spectra derived with the ATLAS~9 and ATLAS~12 
programs \citep{Kurucz:95::} and matched to the observed template spectrum, 
$F(\lambda)$.

\subsection{Iodine cell data pipeline}

In our data pipeline the reduction process involves the following
procedures. An observed stellar spectrum, $I$, is a convolution of an 
intrinsic stellar spectrum, $F$, with a spectrographs' PSF, $\psi$. Such 
a convolution in a discrete form can be written as \citep{Valenti:95::,Endl:00::}
\begin{equation}
I_i = \sum_{j=i-r}^{i+r}F_j\psi_{i-j}, \quad i=1,\dots,N
\end{equation}
where $N$ is the number of pixels in the analyzed spectrum and
$r$ is the range of PSF such as $\psi_l=0$ for $|l| > r$. It is
beneficial to work with an oversampled  version of the above
equation
\begin{equation}
I_i = \sum_{j=oi}^{o(i+1)-1}
\left(\sum_{k=j-r}^{j+r}F_k\psi_{j-k}\right), \quad i=1,\dots,N
\end{equation}
where $o$ is the oversampling factor. This equation can be rewritten
into two other useful formulas
\begin{equation}
I_i = \sum_{j=-r}^{r}
\left(\sum_{k=0}^{o-1}F_{j+k+oi}\right)\psi_{-j}, \quad i=1,\dots,N
\end{equation}
and
\begin{equation}
I_i = \sum_{j=-r}^{r+o-1}
\left(\sum_{k=-j}^{o-j-1}\psi_{k}\right)F_{j+oi}, \quad i=1,\dots,N
\end{equation}
They constitute a set of equations which can be written in the form
${\bf I} = {\mathbb R}{\bf X}$ where ${\bf I}$ represents a vector of
the observed spectrum $I_i, i=1,\dots,N$ and ${\bf X}$ represents 
a vector of the PSF $\psi_l, l=-r,\dots,r$ or of the intrinsic
stellar spectrum $F_k, k=1,...,oN+2r$. The matrix ${\mathbb R}$ can be 
easily inferred from the appropriate sums above. By solving
the equation 5 one can derive the PSF given the observed and 
intrinsic stellar spectrum and by solving the equation 6
one can derive the intrinsic stellar spectrum given the PSF and
the observed spectrum. The latter process is obviously a deconvolution. 
The remaining needed set of equations is one that allows us to clean a stellar 
spectrum observed through an iodine cell from the iodine lines 
as it is symbolically described in the equation 2. This can be 
easily achieved by noting that in such a case $F_k$ in the equation 6 
has to be replaced by $F_k T_k$ where $T_k$ is a known transmission 
function of the iodine cell. In consequence, the set of equations
${\bf I} = {\mathbb R}{\bf X}$ is replaced by ${\bf I} = {\mathbb R}{\mathbb T}{\bf X}$
where ${\mathbb T}$ is a diagonal matrix with the diagonal elements equal
to $T_k$ and then solved for ${\bf X}$ ($X_k = F_k, k=1,...,oN+2r$).
In our data pipeline the equations 5 and 6 are solved with the maximum 
entropy method using a commercially available MEMSYS5 software.
Our data pipeline provides state of the art 1 m~s$^{-1}$ or better
RV precision for single stars (see Fig.~\ref{fig2}).

In practice, the data reduction process is carried out as follows. Each observing 
night, a calibration exposure of a rapidly rotating B star or a quartz lamp is 
taken through an iodine cell. Such a spectrum in principle contains only the iodine 
lines and is used to determine the first wavelength solution and the first estimate 
of the PSF. This is done by taking $F = 1$ in the equation 1 and involves 
solving the equation 5 for $\psi_k$ with $F_k = T_k$. The transmission function
$T$ of the iodine cell as a function of wavelength is obviously known. An 
observing sequence for an SB2 involves taking a pair of exposures one with and one 
without the I$_{2}$ cell --- the instantaneous composite template. The PSF is deconvolved 
from the template to obtain an intrinsic template spectrum, $F_k$, using the 
equation 6. The template is assumed to have a wavelength solution from the
calibration exposure. Such a template is then used to determine the PSF and 
the wavelength solution for the exposure of SB2s taken with the cell. This 
is done by using the equation 1 and involves solving the equation 5 where $F_k$
is replaced by $F_k T_k$. In the final step the new PSF is used in 
the equation 6 to obtain a deconvolved and free of the I$_2$ lines
SB2 spectrum. This final spectrum is ready to be used with TODCOR
as in \cite{Konacki:05::}.

\subsection{Tomographic disentangling of composite spectra}

The disentangling of the spectra of SB2s is a well known problem to which
several more or less restricted solutions exist and are described in a
rich literature. We have decided to follow the tomographic approach to the 
disentangling problem by \cite{Bag:91::} which we 
numerically formulate and solve within the framework of MEMSYS5. 
The idea of the tomographic disentangling is presented in Fig.~\ref{fig3}.
Once the real template spectra for the components of a binary are available, 
it is no longer necessary to use TODCOR as one can simply use real templates 
and the original iodine equation \ref{i2::} by replacing $F(\lambda+\Delta\lambda)$ 
with $S^1(\lambda+\Delta\lambda^1) + \delta_S S^2(\lambda+\Delta\lambda^2)$ where
$S^1,S^2$ represent the two templates, $\Delta\lambda^1, \Delta\lambda^2$ 
their shifts and $\delta_S$ the brightness ratio. Still TODCOR remains an 
important intermediate step to get the first approximate RVs that are 
subsequently used in the disentangling scheme. 

It is worth noting that the composite spectra have to be continuum normalized 
before they are tomographically disentangled and that the tomographic disentangling
does not provide the real brightness ratio but only reproduces the depth of the 
spectral lines with respect to the normalized continuum of the composite spectrum.
Hence $\delta_S$ is not the actual brightness ratio. Successful tomographic 
disentangling can be carried out based on as little as $\sim$10 spectra given 
that they sample different radial velocities of the components.

\subsection{RVs and their errors}

The spectra used to derive RVs were collected with four telescopes and three spectrographs 
the Keck I/Hires, Shane/CAT/Hamspec and TNG/Sarg over the years 
2003-2008. All three instruments are echelle spectrographs which in our
setup provide spectra with a resolution of 67 000, 86 000 and 60,000 respectively 
and are equipped with iodine cells. The observing was carried out as described above.
The typical signal to noise ratios (SNRs) per collapsed
pixel of the collected spectra are $\sim$250 for the Keck I/Hires, $\sim$75-150 
for the TNG/Sarg and $\sim$50-150 for the Shane/CAT/Hamspec. These are for
the composite observed spectra of SB2s. Hence for example a brightness
ratio of 2.3 (HD78418) corresponds to an SNR of $\sim$175 for the primary 
component and an SNR of $\sim$75 for the secondary. The RVs of the primaries 
are thus typically more accurate than of the secondaries. The formal
RV errors were computed from the scatter between echelle orders. As it turns out
these errors are underestimated. In order to obtain a reduced $\chi^2$
equal to 1 for a spectroscopic orbital solution 
and thus conservative estimates of the errors of the least-squares best-fit
parameters, we add an additional error in quadrature to the formal errors
(see \S{4}). The most likely cause of the underestimation of the formal errors, in
addition to the RV jitter of the stars, are subtle imperfections in the 
template component spectra obtained with the tomographic disentangling
and used as a reference for RV computation. Figure~\ref{fig4} demonstrates how
subtle the problem is. The sum of tomographically disentangled component 
spectra provide for an overall better match to observed spectra then a single 
composite {\it observed} spectrum. Still in many cases we do not yet reach a 
photon limited RV precision for SB2s. This problem is being investigated and 
future releases of more accurate RVs from our program are very likely. The 
current RVs used in this paper are listed in Table~1.

\subsection{RV modeling}

In a binary's center of mass coordinate system where the xy-plane is in 
the plane of the sky and the z-axis is directed away from the observer, 
the RVs may be modeled with the following equation
\begin{equation}
\label{rv:01::}
V_{total} = V_R + {\cal V}_{z} + \Delta V_{tides} + \Delta V_{SR\,GR}
\end{equation}
where ${\cal V}_{z}$ is a standard RV variation due to a Keplerian
motion, $\Delta V_{tides}$ is an RV variation due to tidal distortion
of the components, $\Delta V_{SR\,GR}$ is a relativistic contribution
to RVs and $V_R$ is the radial velocity of the center of mass of a binary
(i.e. the gamma velocity). 

\subsubsection{Keplerian motion and the light-time effect}

The standard Keplerian part of the RV variations is as follows.
\begin{equation}
{\bf {\cal V}} = ({\cal V}_{x},{\cal V}_{y},{\cal V}_{z}) =
\frac{a\,n}{r}\left( -a\,\bP\,\sin E + a\,\bQ\,\sqrt{1-e^2}\,\cos E \right)
\end{equation}
\[
n = 2\pi/P, \quad r = a\,\left(1 - e\,\cos E\right), \quad
\bP = \bl\,\cos\omega + \bm\,\sin\omega,
\quad
  \bQ = -\bl\,\sin\omega + {\bf
m}\cos\omega,
\]
\[
  \bl = (\cos\Omega, \sin\Omega, 0), \quad
  \bm = (-\cos i\sin\Omega, \cos i\cos\Omega, \sin i), \quad
\]
In particular, ${\cal V}_{z}$ can also be traditionally expressed as
\begin{equation}
{\cal V}_{z} = K\,\left(\cos(f+\omega) + e\,\cos\omega \right), \quad
K = \frac{a\,n\,\sin i}{\sqrt{1-e^2}}, \quad 
\tan \frac{f}{2} = \sqrt{\frac{1+e}{1-e}}\tan\frac{E}{2},
\end{equation}
where $f$ is the true anomaly and $E=E(t)$ is the eccentric anomaly given by 
the Kepler equation $E - e\sin E = 2\pi(t - T_{\mathrm{p}})/P$, $P$ is the 
orbital period, $a,e,i,\omega,\Omega,T_p$ are the standard Keplerian
elements---the semi-major axis of a component's orbit, the eccentricity, the 
inclination, the longitude of pericenter, the longitude of ascending node 
and the time of pericenter. The spectroscopic orbits 
of the components can be fully described using the two respective RV 
amplitudes $K_{1,2} = a_{1,2}\,n\,\sin i/\sqrt{1-e^2}$, with the longitudes of 
pericenter satisfying the relation $\omega_2 = \omega_1 + \pi$ and the 
remaining orbital parameters being the same. Note that here $a_{1,2}$
refers to orbits of the components with respect to the center of mass 
of the binary.

At the level of precision 1-10 m~s$^{-1}$ it is important to include
the light-time effect within the binary's orbit. We achieve this
by solving the following implicit equation
\begin{multline}
T(t_{1,2}) = t_{1,2} + Z_{1,2}(t_{1,2})/c, \\
Z_{1,2}(t_{1,2}) = r\sin i\,\sin\left(f+\omega_{1,2}\right) 
= \frac{K_{1,2}}{c\,n\,\left(1+e\cos f\right)}\,\left(1-e^2\right)^{3/2}
\sin\left(f+\omega_{1,2}\right),
\end{multline}
for $t_{1,2}$, where $T$ is the observed moment of an RV measurement referred
to SSB (i.e. corrected for the light-time effect within the Solar System),
$t_{1,2}$ is the actual moment and $1,2$ refer to the primary and secondary 
respectively. Subsequently, $t_{1,2}$ is used as the proper argument in the
earlier equations. However, it is approximately true that
\begin{equation}
t_{1,2} \approx T - Z_{1,2}(t_{1,2})/c
\end{equation}
and since $Z_{1,2}(t_{1,2})/c$ is small, it can be easily derived that
to the first order the RV contribution from the light-time effect is
\citep[see also][]{Zuc:07::}
\begin{equation}
\Delta V_{LTE,1,2} = \frac{1}{c}K^2_{1,2}\sin^2\left(f+\omega_{1,2}\right)\,
\left(1 + e\cos f\right)
\end{equation}
and can be used to estimate the amplitudes of $\Delta V_{LTE,1,2}$,
$K^2_{1,2}/c$, which are 2.9 and 3.2 m~s$^{-1}$ for HD78418, 
15.1 and 16.0 m~s$^{-1}$ for HD123999, 4.4 and 6.7 m~s$^{-1}$
for HD160922, 2.9 and 4.6 m~s$^{-1}$ for HD200077 and 
7.8 and 20.2 m~s$^{-1}$ for HD210027, for the primary and 
secondary respectively.

\subsubsection{Tidal effects}

The contributions of the tidally distorted binary components to the
observed RVs were first noted by \cite{Ste:41::}. Such contributions arise
from the non-spherical shapes of the binary's components and their non-uniform 
surface brightness due to predominantly gravitational and limb darkenings.  
The corresponding RV variations were analyzed by \cite{Wil:76::} and 
\cite{Kop:80::}. \cite{Kop:80::} made 
an attempt to derive an analytic approximation while \cite{Wil:76::} used a more 
numerical approach employing the Wilson-Davinney code \citep{Wil:71::,Wil:79::} 
for modeling eclipsing binary light curves to compute the ``tidal" RV contribution. 
Recently, this effect was also analyzed by \cite{Eat:08::}. We follow the approach
by \cite{Wil:76::} and \cite{Eat:08::} where the ``tidal " contribution
to the RVs is modeled with the Wilson-Davinney code. To this end we use 
the light curve modeling program {\tt lc} which is part of the  Wilson-Daviney code
assuming the parameters of our targets as in Table 3. The resulting RV 
variations are shown in Fig.~\ref{fig5}. It should be noted that even though
these effects are quite small they can have a significant qualitative 
effect if ignored by e.g. mimicking a small orbital eccentricity in 
an otherwise circular orbit. 

\subsubsection{Relativistic effects}

The relativistic description of the motion of a binary star from the point of
view of its radial velocities was derived by \cite{Kop:99::}. In our model we 
adopt the following relativistic correction based on the equations 76-86
from \cite{Kop:99::}
\begin{multline}
\Delta V_{SR\,GR} = 
\frac{1}{2c}\left(V_R^2 + V_T^2\right) 
- \frac{1}{c}\left(\frac{1}{2}V_E^2 + U_E\right)
- \frac{V_R}{c}{\bf S}\cdot{\bf V}_G 
- \frac{1}{c}\left(\frac{V_G^2}{2} + U_{SS} - 
\left({\bf S}\cdot{\bf V}_G\right)^2\right) + \\
- \frac{1}{c}{\bf S}\cdot{\bf V}_G\,{\cal V}_{z}
- \frac{1}{c}{\bf S}\cdot{\bf V}_G\,\left({\cal V}_x\,\mu_\delta  + {\cal
V}_y\,\mu_\alpha\right)\left(t-t_0\right) + 
\left({\cal V}_x\,\mu_\delta  + {\cal V}_y\,\mu_\alpha\right)\left(t-t_0\right) + \\
\frac{1}{c}V_R\,{\cal V}_{z} +
\frac{1}{c}V_R\,\left({\cal V}_x\,\mu_\delta  + {\cal
V}_y\,\mu_\alpha\right)\left(t-t_0\right) + 
\frac{1}{c}\left({\cal V}_x\,V_x + {\cal V}_y\,V_y\right) +
\frac{1}{c}\left(\frac{1}{2}{\cal V}^2 + U_{C}\right)
\end{multline}
were $V_R,V_T$ are the radial and tangential velocities of the center
of mass of the binary, $V_E,U_E$ are the velocity magnitude and gravitational
potential of the observer with respect to the geocenter, 
${\bf V}_G, V_G$ are the velocity vector and its magnitude of 
the geocenter with respect to the Solar System barycenter (SSB),
$U_{SS}$ is the gravitational potential at the geocenter due to 
all major bodies in the Solar System, $\mu_\alpha,\mu_\delta$
is the proper motion of the center of mass of the binary
expressed in radians per time unit, $V_x,V_x$ are the components of the
tangential velocity vector such as $V_x = pc\,\mu_\alpha/\kappa, 
V_y = pc\,\mu_\delta/\kappa$ where $\kappa$ is the parallax, $pc$
is one parsec, and $\bf S$ is the unit vector
towards the center of mass of the binary. ${\cal V}_x,{\cal V}_y,{\cal V}_z$ 
as before denote the coordinates of the orbital velocity vector of a given
component, ${\cal V}$ is its magnitude and $U_C$ is the gravitational potential at 
the position of a binary component due to the gravitational field
of its companion. Note that ${\bf S}$ varies slowly due to the
proper motion vector $\bmu$, ${\bf S} = {\bf S} + \bmu\,(t-t_0)$
and the term $\left({\cal V}_x\,\mu_\delta  + {\cal V}_y\,
\mu_\alpha\right)\left(t-t_0\right)$ is not really relativistic
but is obviously due to the varying ${\bf S}$. The necessary quantities 
such as ${\bf V}_G,U_E, U_{SS}, V_T, {\bf S}$ we calculated using the JPL 
ephemerides DE405 and the catalogue positions and proper motions 
of the targets with the help of the {\tt NOVAS} library of astrometric
subroutines by \cite{Kap:89::}. 

The last term in the above equation is the combined effect
of the transverse Doppler effect and the gravitational 
redshift and also the dominant term in the relativistic
correction. It can be easily shown that its periodic
part has the following form 
\begin{equation}
\left(\frac{1}{2}{\cal V}^2 +
U_{C}\right)_{periodic} = \gamma_{1,2} \,\cos f, \quad
\gamma_{1,2} = \frac{G M_{2,1}\left( M_{1,2} + 2 M_{2,1}\right)}{c\,a\left(M_{1} + M_{2}\right)}
\frac{e}{\left(1 - e^2\right)}
\end{equation}
for the primary and secondary respectively where $M_1, M_2$ are the masses
of the components. One can employ the above equation to test GR by using 
$\gamma_{1,2}$ as a free parameter and fitting for it. 
However as noted by \cite{Kop:99::}, it is difficult in practice 
due to the coupling of this effect with the Keplerian part of the model. 
The above equation can be rewritten in the following form
\begin{equation}
\gamma_{1,2} =
\frac{K_{1,2}\left(2K_{1,2}+K_{2,1}\right)}{c\,\sin^2 i}\,e
\end{equation}
and, as has been recently noted by \cite{Zuc:07::}, can 
in principle be used to derive the orbital inclination using just the 
RV measurements.

Three of our targets HD78418, HD123999 and HD200077 have significant
eccentricities and the resulting $\gamma_{1,2}$ are as follows:
4.8 and 5.9 m~s$^{-1}$, 9.7 and 10.1 m~s$^{-1}$, 8.1 and 11.0 m~s$^{-1}$
for the primary and secondary respectively. Only for HD78418, the 
procedure for deriving $\sin^2 i$ proposed by \cite{Zuc:07::} provided 
a value of $0.3832$ ($i=141.8$ deg) close to the real one of $0.2985$ 
($i=146.9$) as determined with the help of PTI astrometric data.
However the error of $\sin^2 i=0.3832$ is 1.8 so this is more of a 
coincidence. The method may yet turn out to be useful given sufficiently 
accurate and/or numerous RV data sets. 

Finally, let us note that our attempts to fit for $\dot{\omega}$ 
which could be either due to the relativistic or
tidal precession \citep{Maz:08::} have not produced any meanigful 
results. In four cases $\dot{\omega}$ was at the level of its 
formal error. In one case (HD78418) $\dot{\omega}$ was technically 
relatively significant (2.4$\pm$0.6$\times$10$^{-4}$ deg/day) but its
inclusion in the model have not improved the best-fit rms.
Hence, the orbital part of the description of the motion 
remains in its classic Newtonian form as in Eq.~8 throughout this
paper.

\section{Visibilities and their modeling}

Interferometers such as PTI typically measure a fringe contrast - a normalized
(by the total power received from a source) amplitude of the coherence
function. The normalized visibility of a binary star can be modeled with
the following equation \citep[see][]{Bod:00b::,Konacki:04::}
\begin{equation}
\label{bin:03::}
V^2_{binary} = \frac{V_1^2 + r^2 V_2^2 + 2 r V_1 V_2
\cos{( 2\pi \bB_{\perp}\cdot\Delta\bs /\lambda)} }{(1 + r)^2}
\end{equation}
where $V_1,V_2$  are the visibilities of the components approximated
with the visibility of a uniform disk of diameter $\theta$
\begin{equation}
\label{disk:02::}
V^2_{i} = \left(\frac{2J_1(\pi\theta B_{\perp}\lambda)}
{\pi\theta B_{\perp}\lambda}\right)^2, i=1,2
\end{equation}
where $B_{\perp} = \norm{\bB_{\perp}}$ is the length of the projected
baseline vector of a two aperture interferometer, $r$ is the brightness 
ratio at the observing wavelength $\lambda$ ($r = P_2/P_1$ where $P_1,P_2$ 
are the total powers of the binary components at the given wavelength) and 
$\Delta\bs = (\Delta\delta,\Delta\alpha)$ is the separation vector between 
the primary and the secondary in the plane tangent to the sky.

The separation vector between the primary and the secondary is given by
the following equations \cite[see e.g.][]{Kov:95::,Kam:67::}
\begin{equation}
\label{e:RtKepler}
   \Delta\bs(t) = \left(\Delta\delta(t),\Delta\alpha(t)\right) =
\kappa a\left(\bP\,(\cos E(t) - e) + \bQ\,\sqrt{1 - e^2}\sin E(t)\right),
\end{equation}
here $a$ is the semi-major axis of the {\it relative} orbit and the first
two coordinates of the vectors $\bP,\bQ$ are used. Note that traditionally 
the North direction is the x-axis and the East direction is the y-axis for 
a coordinate system used in astrometry to model the orbital motion. In our 
figures showing the relative orbits, the x-axis is for the right ascension 
and the y-axis for the declination but the orbital motion is modeled as above. 
Note also that a relative astrometric orbit allows for a few possible solutions 
for the pair of angles $\omega,\Omega$ namely $\omega,\Omega$ as well as
$\omega\pm\pi,\Omega\pm\pi$. The inclusion of a spectroscopic orbit
allows one to determine the real $\omega$ and 
traditionally $\Omega$ is chosen to be less than $\pi$.

PTI provides a square of the normalized visibility amplitude, $V^2$, as one 
can obtain its unbiased estimate \citep{Col:99b::,Moz:91::}.
A normalized visibility of a point source should be by definition equal 
to 1. Since a real instrument is not perfect and does not operate in  
a perfect environment, this is typically not the case and the observed
visibility is underestimated. In consequence each visibility measurement has 
to be calibrated. This is carried out by observing at least one calibrator in 
between the observations of a target. The calibrator is typically a single 
star which diameter is known and its visibility can be very well approximated 
with the equation 17. Then the calibrated visibility of a target is given by 
\citep{Bod:98::,Moz:91::}
\begin{equation}
\label{pra:1::}
V^2_{calibrated} = V^{2}_{measured}/V^2_{sys}
\end{equation}
where
\begin{equation}
\label{pra:1a::}
V^2_{sys} = V^{2}_{cal-measured}/V^2_{cal-expected}
\end{equation}
and $V^{2}_{cal-measured}$ is a measured visibility and
$V^2_{cal-expected}$ is an expected visibility of a calibrator
given by the equation (9).

Visibilities of our targets were extracted from the NASA Exoplanet
Science Institute's (NExSci) database of the PTI measurements. They 
were subsequently calibrated using the excellent tools {\tt getCal}
(ver. 2.10.7) and {\tt wbCalib} (ver. 1.7.4) provided by NExSci. The
data reduction was carried out using the default parameters of {\tt
wbCalib}. The software provides $\{t_i,V^2_i,\sigma_i,\lambda_i,u_i,v_i\}$
where $t_i$ is the time of observation, $V^2_i$ is the calibrated visibility
amplitude squared, $\sigma_i$ its error, $\lambda_i$ is the mean
wavelength for the observation, and $(u_i,v_i)$ are the components of the   
projected baseline vector. We do not list them as they can be easily
obtained using the archival PTI data and the available tools.

\section{Least-squares fitting to the combined RV/V$^2$ data sets}

We determined the best fit orbital parameters of the binaries 
in the standard way by minimizing the following function:
\begin{equation}
\chi^2 = \sum_{i=1}^N (V_i^2 - \widehat{V}_i^2)^2/\sigma_{V^2_i}^2
+ \sum_{i=1}^M (RV1_i - \widehat{RV1}_i)^2/\sigma_{RV1_i}^2
+ \sum_{i=1}^M (RV2_i - \widehat{RV2}_i)^2/\sigma_{RV2_i}^2
\end{equation}
were $V^2_i,\sigma_{V^2_i}$, $RV1_i,\sigma_{RV1_i}$ and
$RV2_i,\sigma_{RV2_i}$ are respectively visibilities
and their errors, radial velocities of the primary and their
errors and radial velocities of the secondary and their
errors; $N$ and $M$ denote the number of measurements. 
The symbols with a hat denote the respective model
values computed as described in the previous sections.
The fit was carried out using the Levenberg-Marquardt 
algorithm for solving the minimization problem which was incorporated
into our own software for modeling and least-squares fitting
of the visibilities and radial velocities. As mentioned above,
the RV variations due to the tidal distortions of the components
were modeled with the Wilson-Davinney's {\tt lc} code 
(version from 2007) which was also incorporated into our 
software.

The least-squares fitting formalism allows one to compute the formal uncertainties 
(errors) of the best fit parameters. However such errors do not neccesarily 
correspond to the true uncertainties of the parameters. In order to provide
conservative estimates of these errors we also accounted for the systematic 
errors in the modeling which are due to (1) the uncertainty in the projected 
baseline $(u,v)$ (2) the uncertainty in the mean wavelength $\lambda$ (3) the 
uncertainties in the diameters, $\theta_j$, of the calibrators (which affect 
the modeled visibility through the equation 20) and the diameters of the binary 
components (which affect the modeled visibility through the equation 17) and (4) 
the uncertainties in the parameters used to model the tidal RV effects.
We assumed the following estimates for these uncertainties (1) 0.01 percent
in $(u,v)$, (2) 0.5 percent in $\lambda$ and (3) 10 percent in the calibrator 
and binary components diameters and (4) 10 percent in all the parameters
from Table 2 except for the temperatures for which we assumed an uncertainty
of 1 percent for HD210027 and 2 percent for the other targets; and
for the metallicities we assumed an uncertainty of 0.05 dex. The errors of 
the parameters shown in Tables 4-6 include the contribution from the 
systematic errors.

Our procedure of deriving RVs results in slightly different zero-points
of the velocities for each data subset. These small differences are among
others due to a mismatch, different for each component of a binary, of the synthetic 
templates and the real spectra. The synthetic templates, as already explained, 
are used in the first step of RV computations and the shifts are then carried 
over to the real disentangled component spectra. The other sources of the small 
zero-point differences are due to the use of different spectrographs
and the accuracies with which the zero points of the template iodine cell
transmission functions are known. The velocity shifts are determined
together with the other parameters in the least-squares fits and are
shown in Table 5. Even though the total formal errors of the gamma velocities
are small, due to the above reasons it is unlikely that their realistic
uncertainties are smaller than $\sim$$0.1$ km~s$^{-1}$. 

The full statistical detailes of the best-fit solutions are given in Table 5.
For the RV data sets we determined the additional RV errors added in the
quadrature to the formal RV errors by performing separate orbital fits
to the RVs alone and finding such additional RV errors which provide
$\chi^2 \approx 1$. For $V^2$ measurements we used the errors as
computed with {\tt wbCalib}. However by performing a preliminary fit to 
all the available $V^2$s and RVs, we found those $V^2$ measurements that
significantly deviate from the best-fit solutions. Such deviant
measurements are most likley due to poor weather conditions and/or
an errant behaviour of the instrument. In the final fit we did not use them. 
In Figures 7,9,11,13,15 they are denoted with open (red) circles. The 
resulting total $\chi^2$ values are close or very close to $1$ which 
ensures one that the estimates of the errors of the parameters are realistic.

\subsection{Notes on the individual solutions}

The orbital parameters of the combined spectroscopic-astrometric solutions
are shown in Tables~4-5 and the resulting physical parameters are shown
in Table~6. The masses expressed in the solar masses
were derived with GM$_{\odot}=1.3271244017987\times10^{20}$ m$^{3}$s$^{2}$,
a value used in the DE405 JPL ephemerides. Let us note that the official
value of G recommended by CODATA (The Committee on Data for Science and
Technology) is $6.67428\pm0.00067\times10^{-11}$ m$^{3}$kg$^{-1}$s$^{-2}$.
Hence G is known with a fractional error of 0.01\% and in consequence the
mass of the Sun and the masses of stars expressed in absolute units are known
with only such precision.

We also calculated limits to circumbinary planets for three new
systems HD123999, HD160922 and HD200077. They were calculated as in
\cite{Konacki:09b::} and are shown in Fig.~18. 

{\em HD 78418 (75 Cnc, HR 3626, HIP 44892; V = 5.98 mag, K = 4.37 mag)} 
is a $\sim$19.4 day period binary with a spectral type G5IV-V. The most
recent spectroscopic solution is by \cite{Medeiros:99::} and is
characterized by an RV rms of 461 m~s$^{-1}$. Our combined RV solution
has an rms of 13.1 and 24.3 m~s$^{-1}$ for the primary and secondary
respectively. The binary was resolved with PTI by \cite{Lane:99::} but a
detailed spectroscopic-astrometric analysis was never published. Using 
the PTI data archive, we extracted 440 V$^2$ measurements spanning 1999-2008 
of which 415 were used in the final fit by adopting a cutoff at 0.1 for V$^2$'s 
O-C. The masses of the components are 1.173$\pm$0.024 M$_{\odot}$ (2.0\% accuracy) 
and 1.011$\pm$0.021 M$_{\odot}$ (2.1\% accuracy) for the primary and secondary
respectively. This is a far more accurate mass determination than the one
from \cite{Lane:99::}. The masses and the absolute magnitudes in the K and H 
bands together with the Padova isochrones \citep{Mar:08::} 
allow us to estimate the age of the system as about 2.5-4.0 Gyr. 
For the isochrones we adopted a metallicity of -0.09
\citep{Nord:08::,Nord:04::}. The JHK photometry comes from the 2MASS
catalogue \citep{Scr:06::}.

{\em HD 123999 (12 Boo, HR 5304, HIP 69226; V = 4.82 mag, K = 3.64)} is 
a $\sim$9.6 day period binary with spectral type F8IV. The latest spectroscopic 
orbit was published by \cite{Tom:06::}. It has an RV rms of 180 and 100 m~s$^{-1}$
for the primary and secondary respectively. Our combined RV data sets 
are characterized by an rms of 34.0 and 38.3 m~s$^{-1}$ which is 
due to somewhat wide spectral lines of the stars. The binary was resolved
and studied with PTI by \cite{Bod:00a::} and most recently by \cite{Bod:05::} 
who obtained a combined spectroscopic-astrometric solution using the PTI V$^2$ 
measurements collected in the years 1998-2004. We extracted 374 V$^2$ measurements
spanning 1999-2008 and used 346 in the final fit (a cutoff at 0.09 for V$^2$'s O-C).
The masses of the components are 1.4109$\pm$0.0028 M$_{\odot}$ (0.2\% accuracy) 
and 1.3677$\pm$0.0028 M$_{\odot}$ (0.2\% accuracy) for the primary and secondary
respectively. These values are consistent with those by \cite{Bod:05::} 
but about two times more accurate. The RVs used by \cite{Bod:05::} were
characterized by an rms of 400 and 490 m~s$^{-1}$. Assuming a solar 
metallicity as \cite{Bod:05::}, the estimated age for the system is 2.5-2.9 Gyr.
Note that as in \cite{Bod:05::}, the secondary appears to be younger than
the primary. Our smaller errorbars for the masses and absolute K and H band
magnitudes make this discrepancy even more apparent.
The JHK photometry comes from \cite{Bod:00a::}.

{\em HD 160922 ($\omega$ Dra, HR 6596, HIP 86201; V = 4.90 mag, K = 3.62 mag )} is a
$\sim$5.3 day period binary with a spectral type F5V. An improved
spectroscopic solution was recently published by \cite{Fe:09::}.
It is characterized by an RV rms of 190 m~s$^{-1}$. Our combined RV solution  
has an rms of 37.9 and 31.9 m~s$^{-1}$ for the primary and secondary
respectively. This somewhat lower RV precision is due to relatively wide
spectral lines of the components. The PTI archive provides 62 never 
published V$^2$ measurements of which 61 were used in the final fit 
(V$^2$'s O-C cutoff at 0.08). We have obtained a spectroscopic-astrometric 
orbital solution with a small but statistically significant 
eccentricity (0.00220$\pm$0.00031) which is in agreement with 
\cite{Fe:09::}. However they have decided to adopt a circular solution 
afterall. The masses of the components are 1.46$\pm$0.16 M$_{\odot}$ (11\% accuracy) 
and 1.18$\pm$0.13 M$_{\odot}$ (11\% accuracy) for the primary and secondary
respectively. The masses are derived for the first time for this system
but their accuracy is poor due to a limited number of V$^2$ measurements
and the near face-on orbital configuration. Based on the derived masses and 
the K band absolute magnitudes we estimate the age of the system as 0.004-2.5
Gyr; a wide range due to large error bars both in the masses and the K-band
magnitudes. For the isochrones we adopted a solar metallicity 
\citep{Nord:08::,Nord:04::}. The JHK photometry comes from the 2MASS
catalogue.

{\em HD 200077 (HIP 103641; V = 6.57 mag, K = 5.12 mag)} is a $\sim$112.5 day period binary 
with a spectral type G0V. The spectrscopic orbital solution for both
components was for the first time derived by \cite{Gold:02::}. Their RVs are
characterized by an rms of 600 and 2150 m~s$^{-1}$ for the primary
and secondary respectively.  Our combined RV solution has an rms of 
respectively 8.5 and 30.4 m~s$^{-1}$. The binary was resolved with PTI 
but this has never been published before. The PTI archive provides 395 
V$^2$ measurements spanning 2001-2007 of which 329 were used in the 
final fit (V$^2$'s O-C cutoff at 0.09). The masses of the components
derived for the first time are 1.1860$\pm$0.0057 M$_{\odot}$ (0.48\% accuracy) 
and 0.9407$\pm$0.0049 M$_{\odot}$ (0.52\% accuracy) for the primary and secondary
respectively. Based on the masses, K and H band absolute magnitudes the 
estimated age is 2.9-4.3 Gyr. For the isochrones we adopted a metallicity 
of -0.14 \citep{Nord:08::,Nord:04::}. The JHK photometry comes from the 2MASS
catalogue.

{\em HD 210027 ($\iota$ Peg, 24 Peg, HR 8430, HIP 109176; V = 3.76 mag, K =
2.56)} is a $\sim$10.2 day period binary with spectral types F5V/G8V. It is one of 
the first SB2s resolved with PTI. The most
recent orbital solution is a combined spectroscopic-astrometric solution
by \cite{Boden:99::}. It is based on the PTI V$^2$ measurements from 1997
and the archival RV measurements from \cite{Fek:83::} 
which are characterized by an rms of about 600 and 700 m~s$^{-1}$ for the
primary and secondary.  Our combined RV
solution has an rms of respectively 17.1 and 69.7 m~s$^{-1}$. The PTI archive
provides 299 V$^2$ measurements spanning 1997-2006 of which 266 were used in the   
final fit (V$^2$'s O-C cutoff at 0.088). The masses of the components  
are 1.33249$\pm$0.00086 M$_{\odot}$ (0.065\% accuracy) and 0.83050$\pm$0.00055 
M$_{\odot}$ (0.066\% accuracy) for the primary and secondary respectively.
This is about a 20 times more accurate mass determination than the one from
\cite{Boden:99::} and constitues the most accurate mass measurement
for a normal star. Based on the masses, K and H band absolute magnitudes 
the estimated age is 4-663 Myr which i sin agreement with \cite{Mo:00::}.
For the isochrones we adopted a solar metallicity \citep{Mo:00::}. 
The JHK photometry comes from \cite{Bou:91::}. Note that our orbital
solution has a small but statistically significant eccentricity 
(0.001764$\pm$0.000063). This combined with the systems' young age may
prove usefull in studying the tidal history of HD~210027.


\section{Discussion}

There are a few ways to determine accurate masses of normal stars. Perhaps
the most classic one is through absolute astrometric orbits of both components
of a binary. However this constitues a challenging measurement and in practice
masses are typically derived using diverse data sets e.g. (1) by
combining RVs, relative astrometry and parallax for single-lined 
spectroscopic binaries, (2) as in this paper by combining relative 
astrometry and RVs for double-lined spectroscopic binaries and (3)
by combining light curves and RVs for eclipsing double-lined
binaries. The last method is the most useful one as it not only
provides the most accurate masses due to a convenient edge-on geometry
(note that the masses are derived from their respective M $\sin^3i$)
but also enables one to determine the radii of stars.

In a recent review, \cite{Torres:09::} collect 118 detached binary stars 
(including 94 eclipsing) with the most accurate mass determinations in the 
literature. These are denoted with open circles in Fig.~18. Even though our 
targets are not eclipsing and the orbital inclinations and their errors
have a significant impact on the precision in masses, 
for two stars HD123999 and HD210027 we have obtained more accurate
mass determinations than for any of the stars from \cite{Torres:09::}.
The accuracies of the masses of these two binaries are in the
precision range covered by only close double neutron star systems 
characterized with radio pulsar
timing. In particular, the masses for HD210027 rival in precision 
the mass determination of the components of the relativistic double 
pulsar system PSR~J0737-3039 \citep{Nice:08::}. If our targets were 
all eclipsing and the accuracy in masses was limited by our RVs alone,
the accuracy would be in the range 0.02\% to 0.42\% (the fractional
accuracy of M $\sin^3i$). The lower limit of this range is equal
to the mass accuracy of PSR~B1913+16 which has the most accurate 
mass determination for a body outside the Solar System \citep{Nice:08::}. 
In fact if we adopt our RV precision and use the achievable from the ground
precision in the orbital inclination angle for eclipsing binaries, we 
can expect to obtain masses with a fractional precision of $\sim$0.001\%
(see Fig.~18). Clearly, our RV technique for double-lined 
and eclipsing binary stars opens an exciting opportunity for derving
masses of stars (and other parameters) with an unprecedented precision.
These combined with parallax measurements from e.g. the planned GAIA 
astrometric mission and hopefully accurate abundance determinations 
should produce an outstanding set of parameters to tests models of
the stellar structure and evolution.

\acknowledgments

We thank the California and Carnegie Exoplanet Search team,
and Geoff Marcy in particular, for allowing us access to their precision
velocimetry tools at Lick Observatory. 
This work benefits from the efforts of the PTI collaboration members who have
each contributed to the development of an extremely reliable observational
instrument. We thank PTI's night assistant Kevin
Rykoski for his efforts to maintain PTI in excellent condition and operating
PTI. This research has made use of the Simbad 
database, operated at CDS, Strasbourg, France.  MWM acknowledges support from the 
Townes Fellowship Program. M.K. is supported by the Foundation for Polish Science 
through a FOCUS grant and fellowship and by the Polish Ministry of Science and Higher 
Education through grant N203 005 32/0449. Part of the
algorithms used in this analysis were developed during the SIM Double Blind 
Test, under JPL contract 1336910. This research has made use of the Simbad 
database, operated at CDS, Strasbourg, France. The observations on the
TNG/SARG have been funded by the Optical Infrared Coordination network (OPTICON),
a major international collaboration supported by the Research Infrastructures 
Programme of the European Commissions Sixth Framework Programme.
This publication makes use of data products from the Two Micron All Sky
Survey, which is a joint project of the University of Massachusetts and the
Infrared Processing and Analysis Center/California Institute of Technology,
funded by the National Aeronautics and Space Administration and the National
Science Foundation.

{\it Facilities:} \facility{Keck I/Hires, TNG/Sarg, Shane/Hamspec, PTI}

\clearpage

%
%

\begin{figure}[ht]
\begin{center}
\includegraphics[width=5.3in]{fig1.ps} 
\caption{Radial velocity precision (an rms from the orbital fit) for the primaries 
of double-lined spectroscopic 
binaries as a function of the publication date based on the ninth catalogue of
spectroscopic binary orbits \citep{Pour:04::}. It is worth noting that 
already in the early XX century an RV precision of several km~s$^{-1}$ for 
double-lined binaries was possible \citep[HD83808]{Plum:08::}.
Current and previous precision range from our method is delimited with 
the red stars (note the logarithmic scale).}
\label{fig1}
\end{center}
\end{figure}

%
%

\begin{figure}
\epsscale{0.8}
\plotone{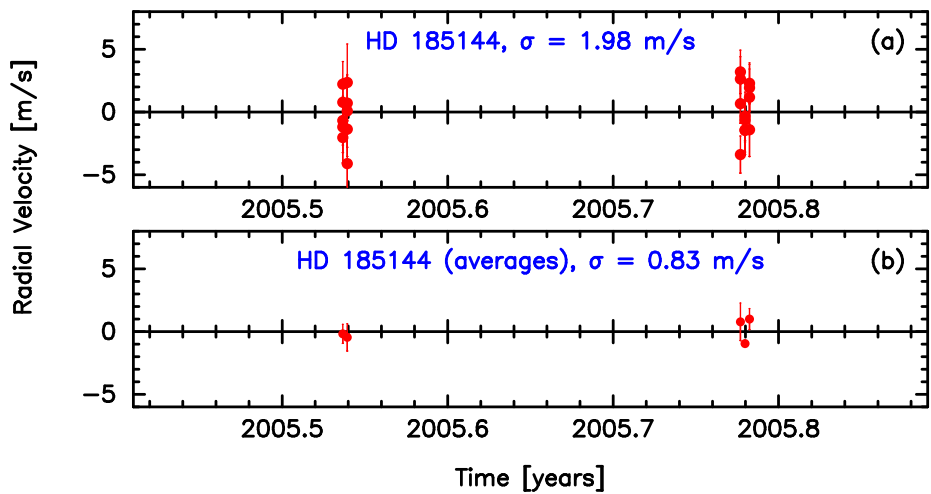}
 \caption{A series of the precision RV measurements of a single
star HD185144 (V=4.7 mag, K0V) taken with the Keck I/HIRES and reduced
using our iodine cell data pipeline. The data were taken on 5 nights:
two in July (5 consecutive spectra each night) and three in October 2005 
(four consecutive spectra each night). The average SNR was 600 for
the template exposure as well as for all the spectra taken with the
iodine cell.}
\label{fig2}
\end{figure}

%
%

\begin{figure}[ht]
\begin{center}
\includegraphics[width=5.3in]{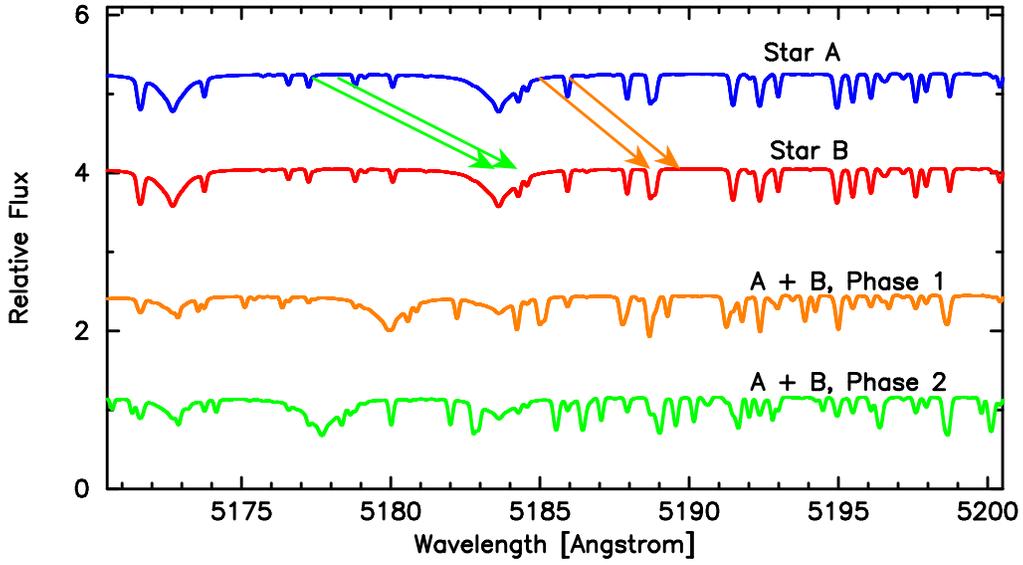} 
\caption{An SB2 as nature's realization of the tomographic imaging.
The composite spectrum of an SB2 taken at any orbital phase
can be interpreted as a result of imaging of a two-layered object.
By observing an SB2 at several different orbital phases and hence
different RVs of its components, one can carry out an equivalent 
of tomographic imaging and eventually be able to disentangle 
the component (``layer") spectra.}
\label{fig3}
\end{center}
\end{figure}

%
%

\begin{figure}[ht]
\begin{center}
\plottwo{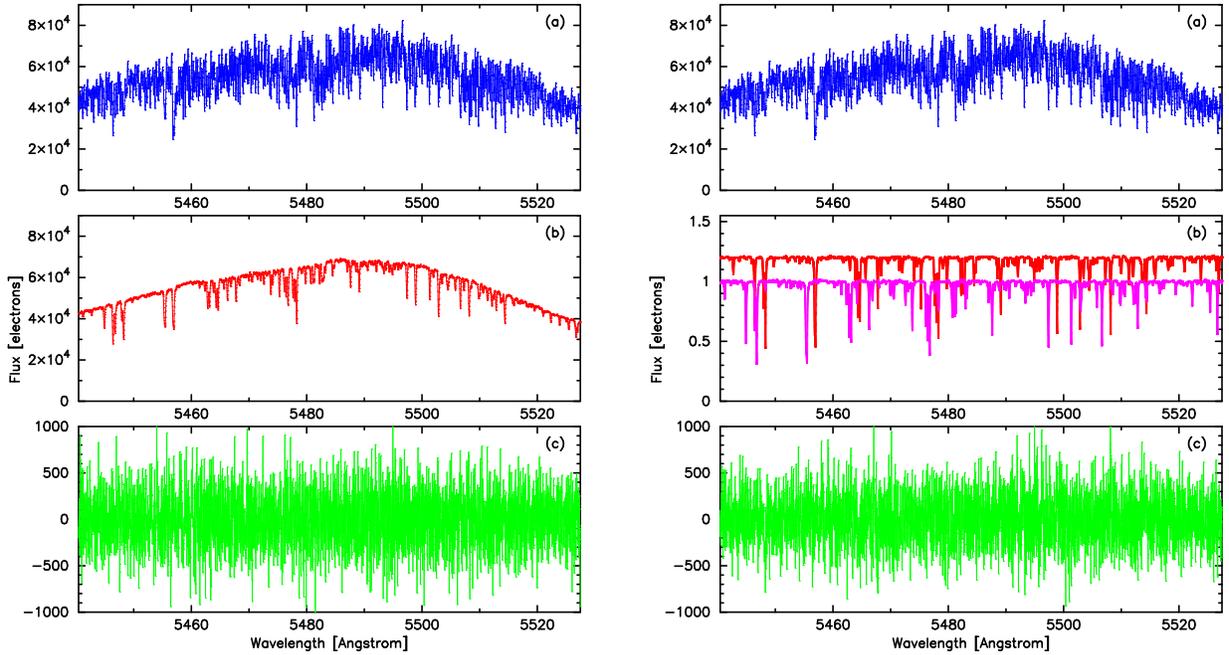}{fig4b.ps}
\caption{Left and right {\it (a)}: a piece of a spectrum of HD4676 as a function 
of wavelength as seen through an iodine cell. Left {\it (b)}: a template
composite spectrum of HD4676 taken just after the iodine cell exposure
shown in {\it (a)}. Right {\it (b)}: the corresponding continuum normalized
disentangled component spectra of HD4676 shifted accordingly to their respective 
RVs. Left and right {\it (c)}: the residuals after subtracting {\it (b)} 
from {\it (a)} and accounting for the iodine cell lines. The residuals from 
the disentangled component spectra are smaller than from the observed composite 
spectrum.}
\label{fig4}
\end{center}
\end{figure}

%
%

\begin{figure}
\figurenum{5}
\epsscale{0.8}
\plotone{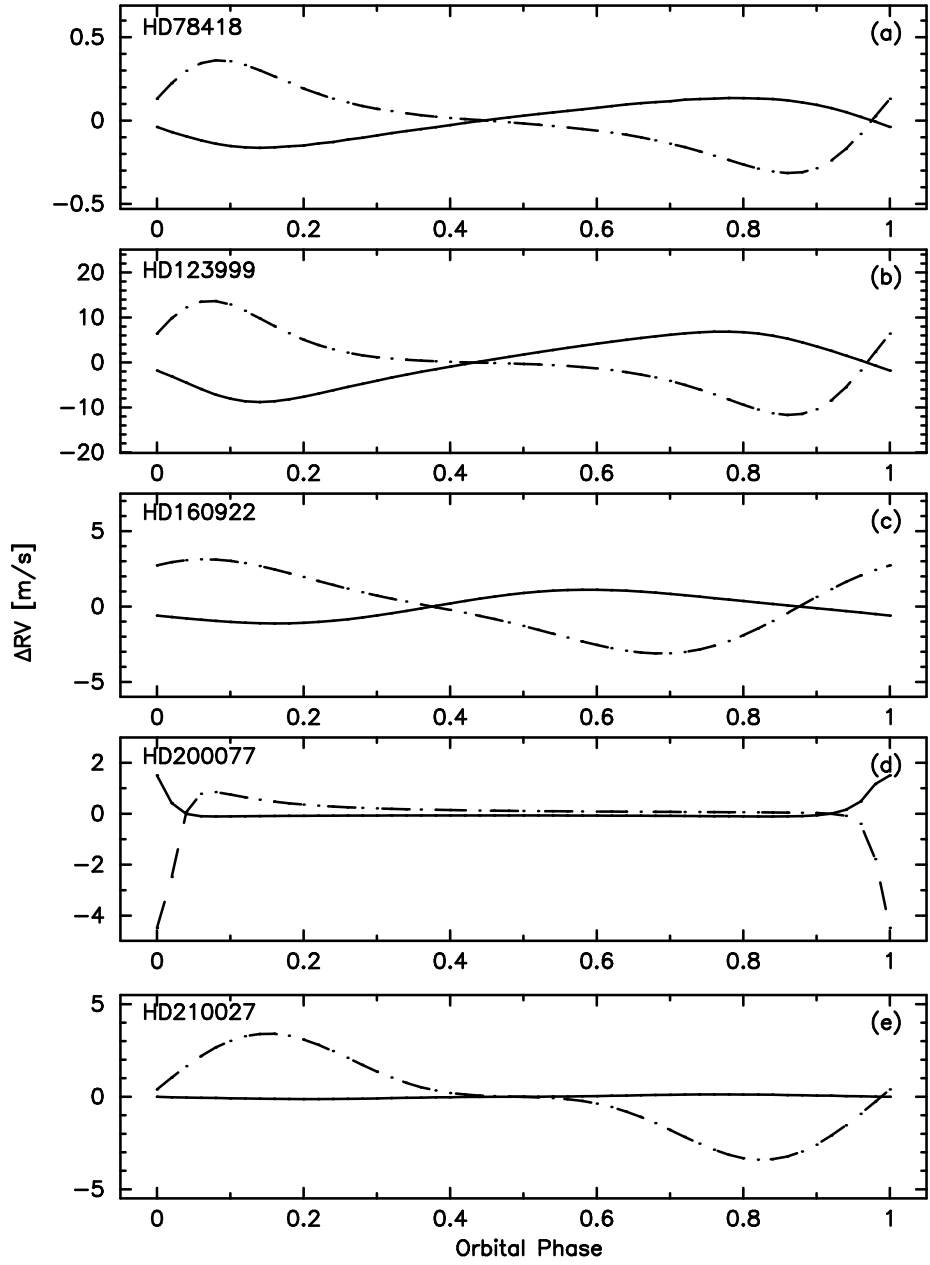}
\caption{Radial velocity variations due to tidal distortion of the binary
components as a function of the orbital phase for HD78418 (a), HD123999 (b),
HD160922(c), HD200077 (d) and HD210027 (e). The solid line is for
the primary and the dashed-dotted line for the secondary.}
\label{fig5}
\end{figure}

%
%

\begin{figure}
\figurenum{6}
\epsscale{0.8}
\plotone{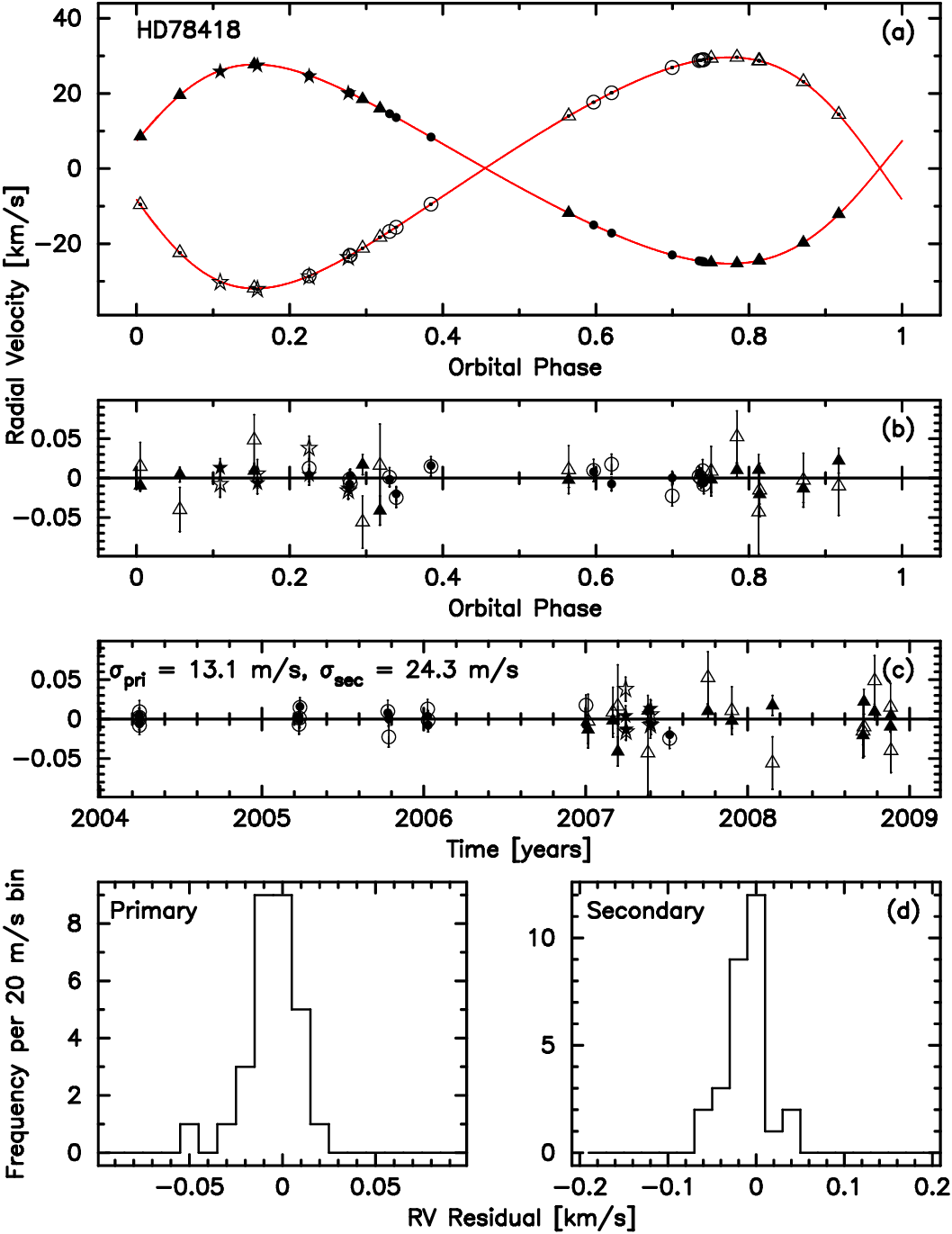}
\caption{Observed and modelled radial velocities of HD78418 as a
function of the orbital phase (a), their best-fit residuals as
a function of the orbital phase (b) and time (c). The histograms
of the residuals for the primary and secondary (d). The Keck I/Hires is
denoted with circles, Shane/CAT/Hamspec with triangles and TNG/Sarg with stars.}
\end{figure}

%
%

\begin{figure}
\figurenum{7}
\epsscale{0.8}
\plotone{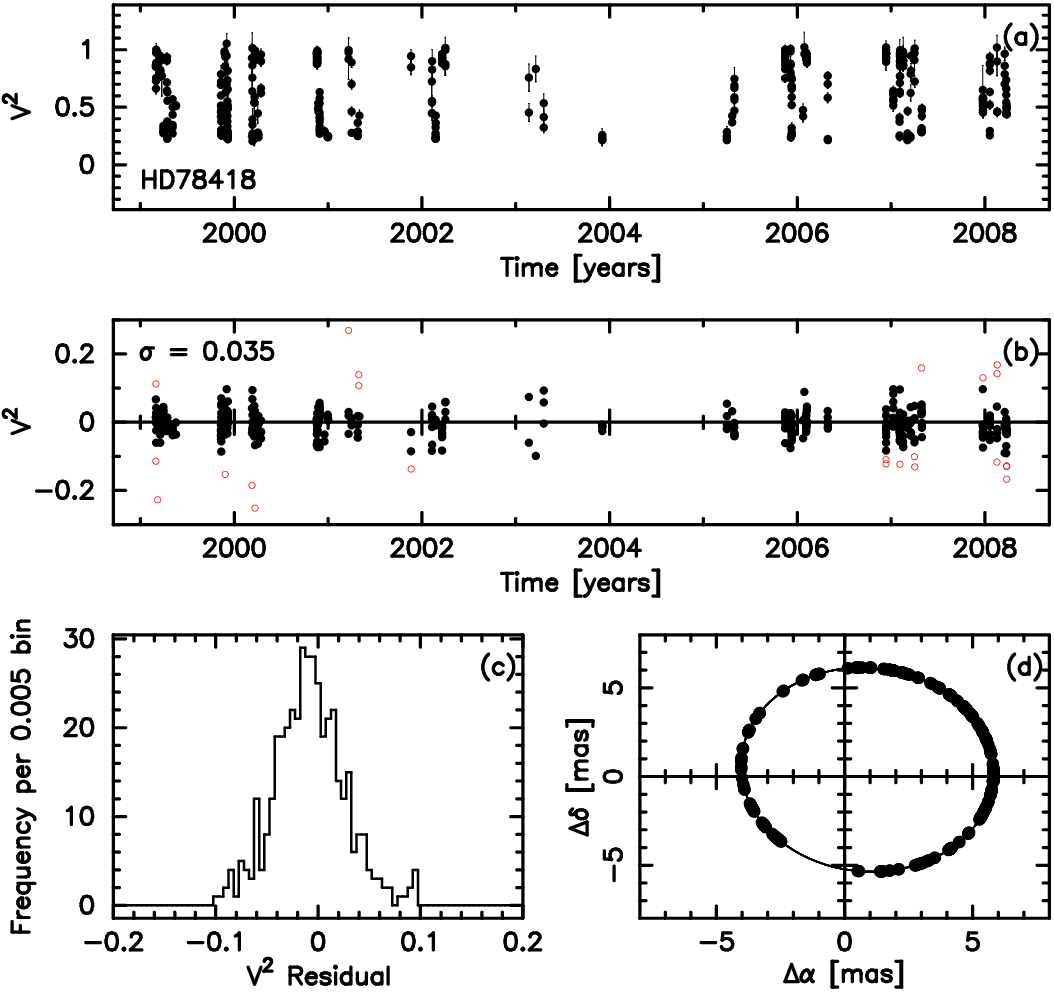}
\caption{Visibility measurements of HD78418 as a function of time (a),
their best-fit residuals as a function of time (b) and histogram (c).
The measurements used to determine the best-fit orbital solution
are denoted with filled circles. The corresponding orbital coverage 
and the relative orbit is shown in the panel (d).}
\end{figure}

%
%

\begin{figure}
\figurenum{8}
\epsscale{0.8}
\plotone{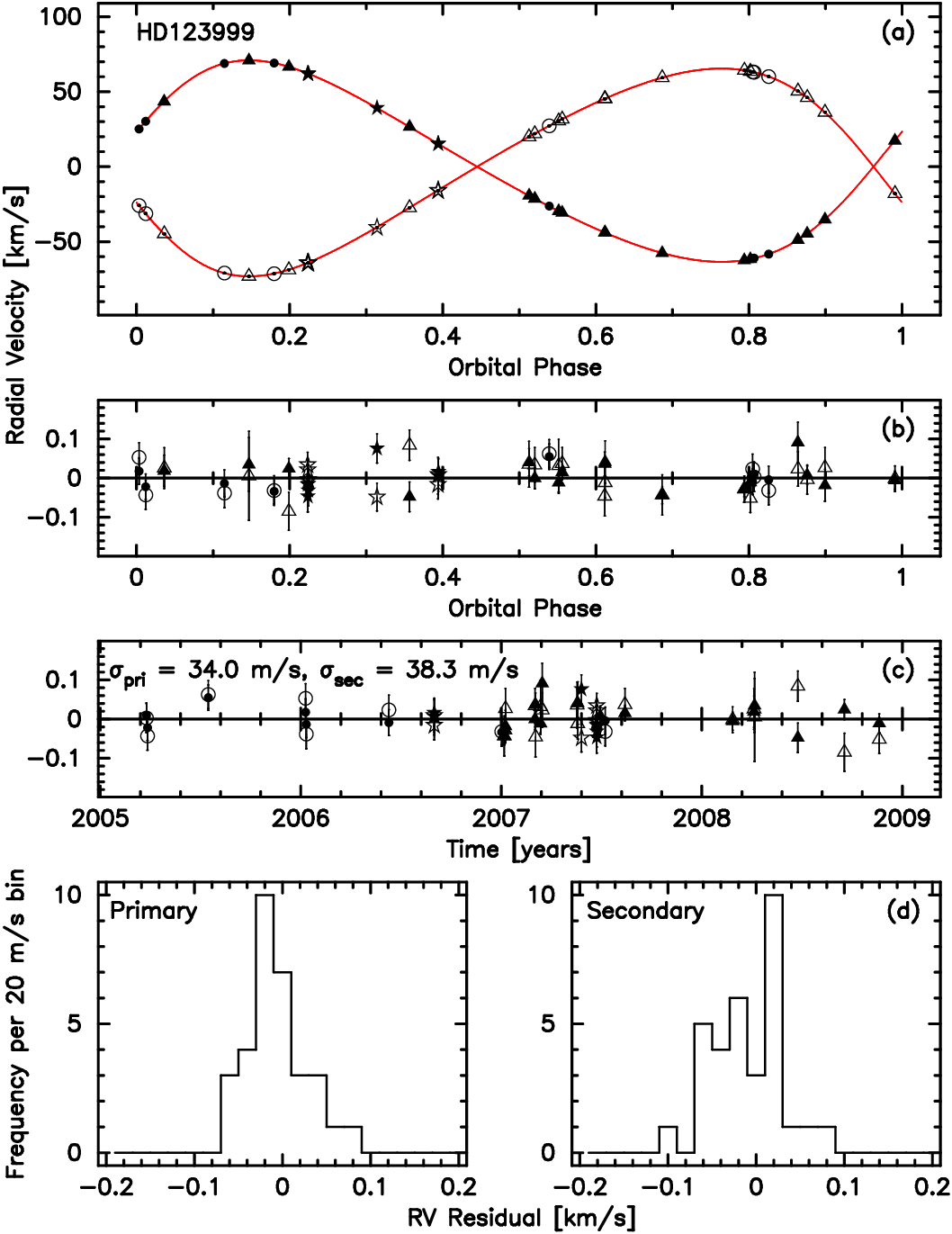}
\caption{Observed and modelled radial velocities of HD123999 as a
function of the orbital phase (a), their best-fit residuals as
a function of the orbital phase (b) and time (c). The histograms
of the residuals for the primary and secondary (d). The Keck I/Hires is
denoted with circles, Shane/CAT/Hamspec with triangles and TNG/Sarg with stars.}
\end{figure}

%
%

\begin{figure}
\figurenum{9}
\epsscale{0.8}
\plotone{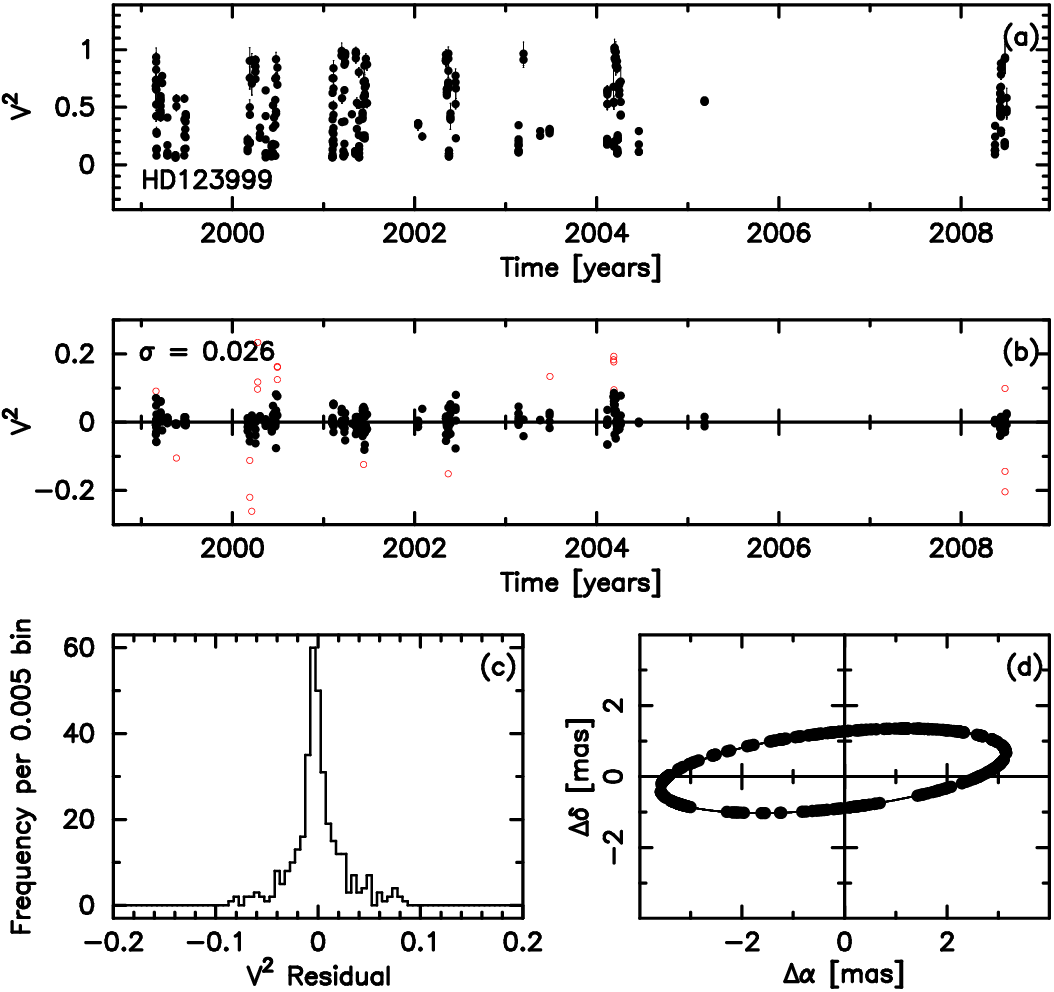}
\caption{Visibility measurements of HD123999 as a function of time (a),
their best-fit residuals as a function of time (b) and histogram (c).
The measurements used to determine the best-fit orbital solution
are denoted with filled circles. The corresponding orbital coverage 
and the relative orbit is shown in the panel (d).}
\end{figure}

%
%

\begin{figure}
\figurenum{10}
\epsscale{0.8}
\plotone{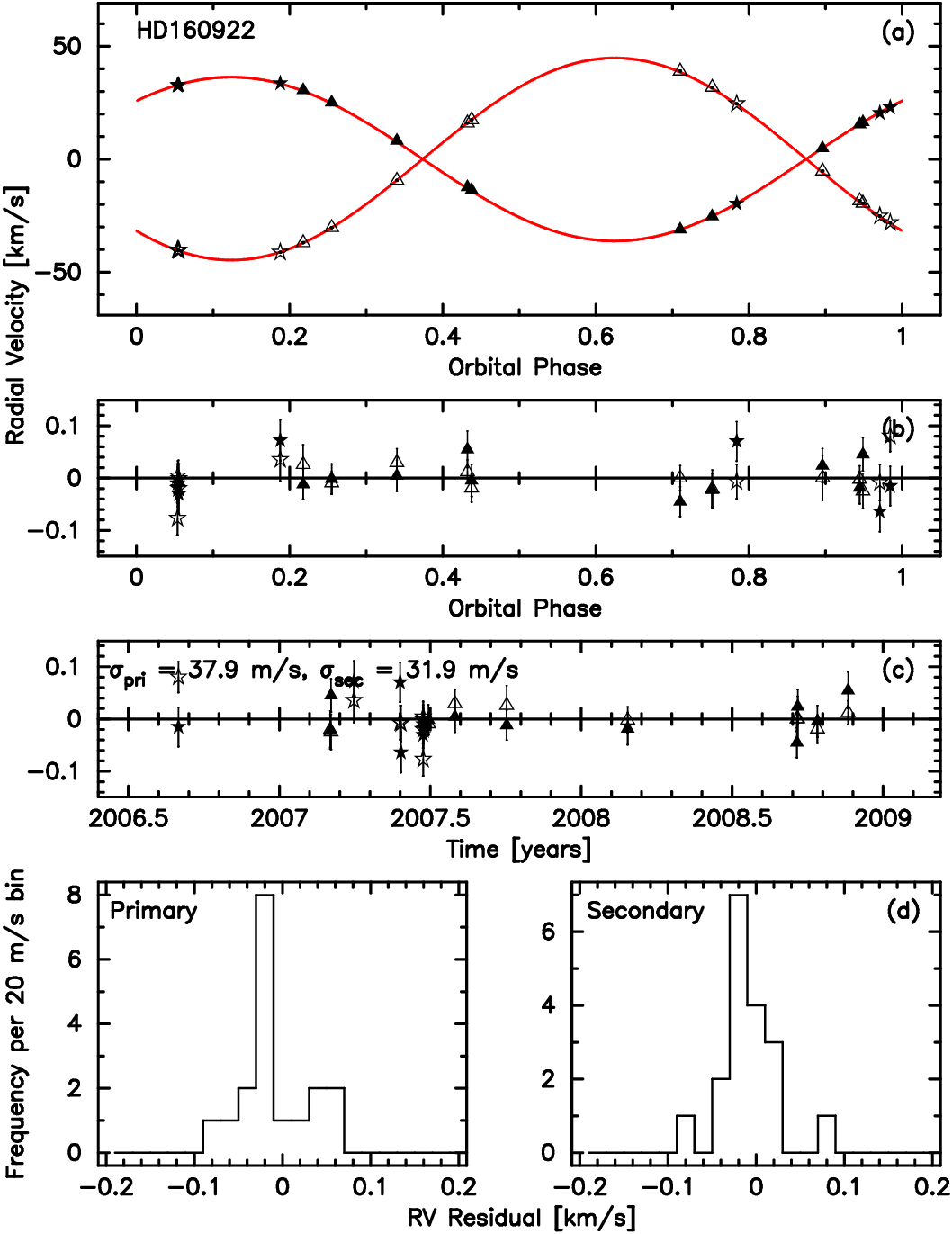}
\caption{Observed and modelled radial velocities of HD160922 as a
function of the orbital phase (a), their best-fit residuals as
a function of the orbital phase (b) and time (c). The histograms
of the residuals for the primary and secondary (d). The Keck I/Hires is
denoted with circles, Shane/CAT/Hamspec with triangles and TNG/Sarg with stars.}
\end{figure}

%
%

\begin{figure}
\figurenum{11}
\epsscale{0.8}
\plotone{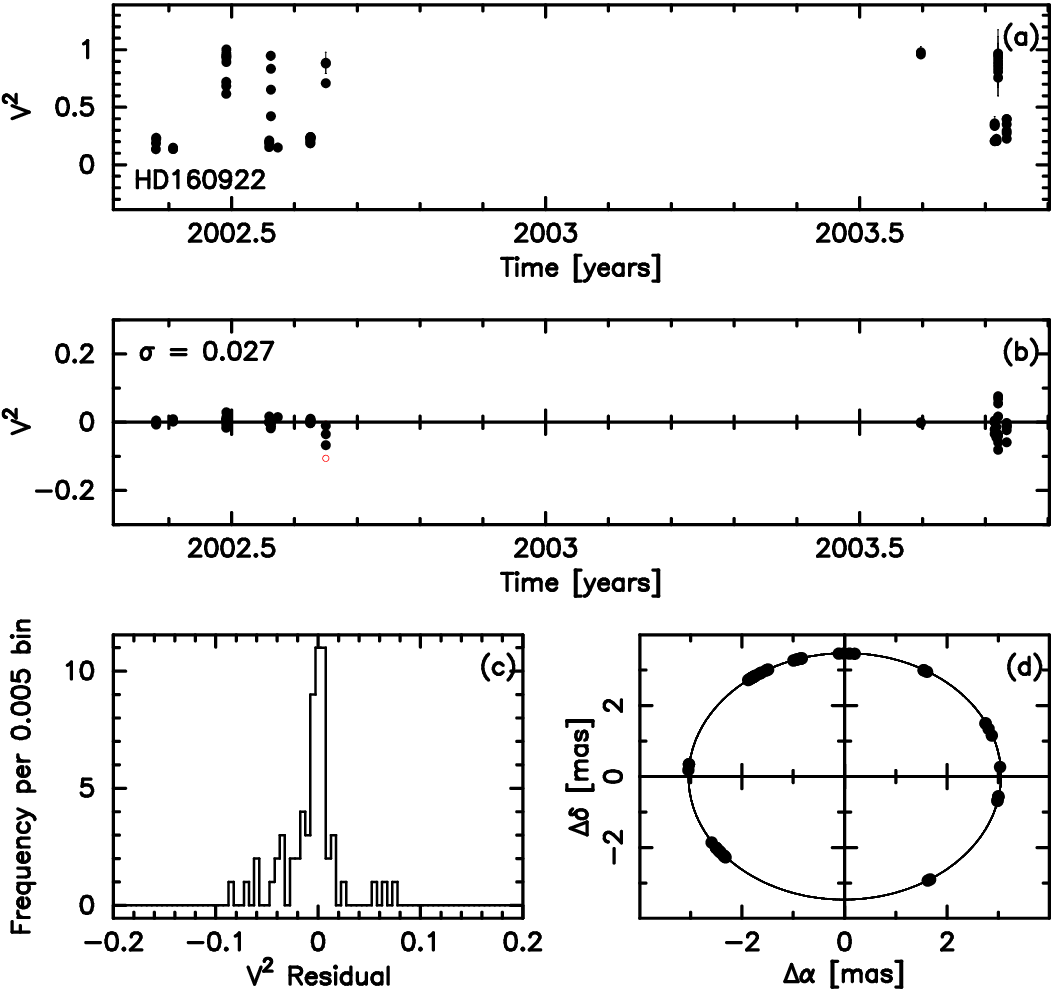}
\caption{Visibility measurements of HD160922 as a function of time (a),
their best-fit residuals as a function of time (b) and histogram (c).
The measurements used to determine the best-fit orbital solution
are denoted with filled circles. The corresponding orbital coverage 
and the relative orbit is shown in the panel (d).}
\end{figure}

%
%

\begin{figure}
\figurenum{12}
\epsscale{0.8}
\plotone{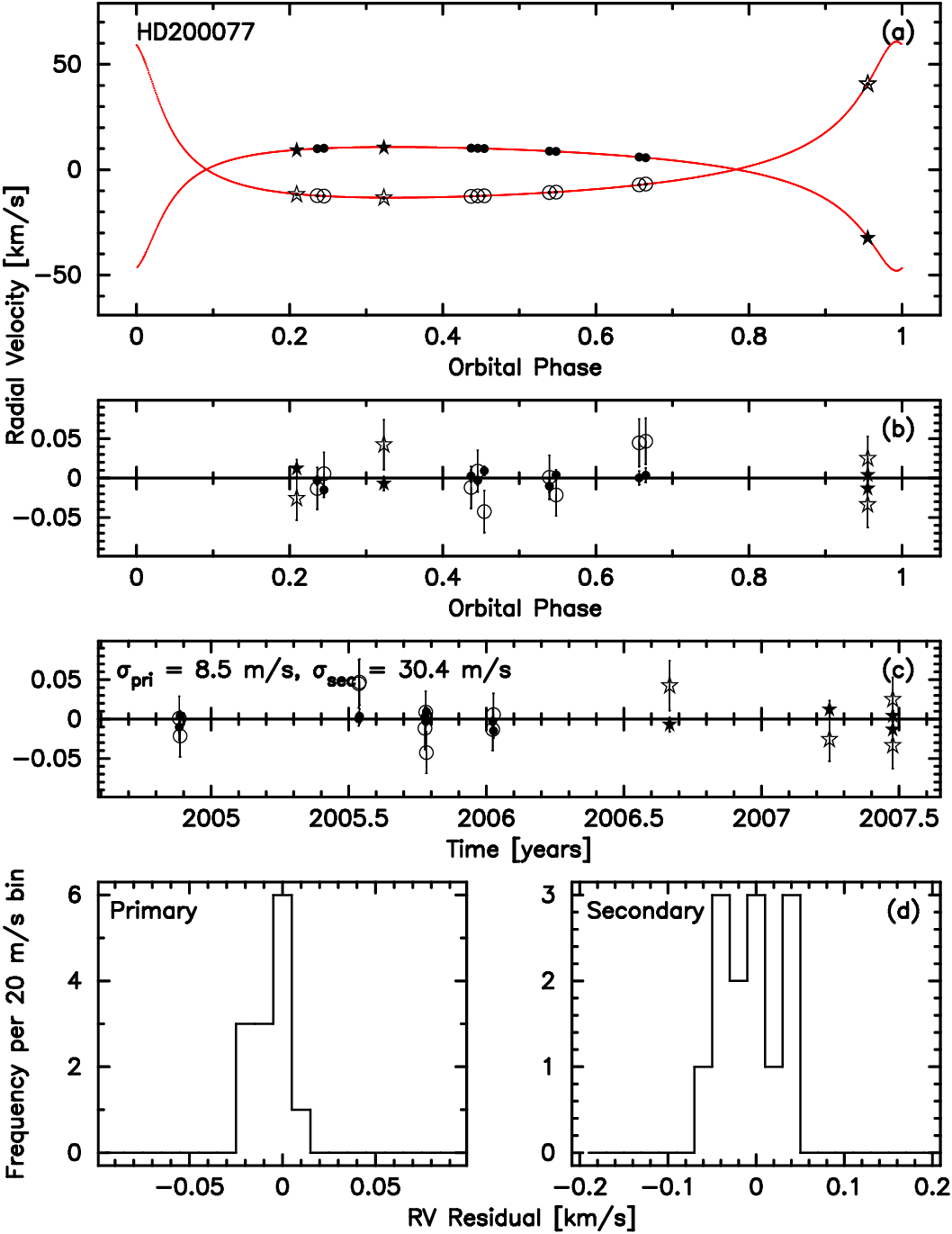}
\caption{Observed and modelled radial velocities of HD200077 as a
function of the orbital phase (a), their best-fit residuals as
a function of the orbital phase (b) and time (c). The histograms
of the residuals for the primary and secondary (d). The Keck I/Hires is
denoted with circles, Shane/CAT/Hamspec with triangles and TNG/Sarg with stars.}
\end{figure}

%
%

\begin{figure}
\figurenum{13}
\epsscale{0.8}
\plotone{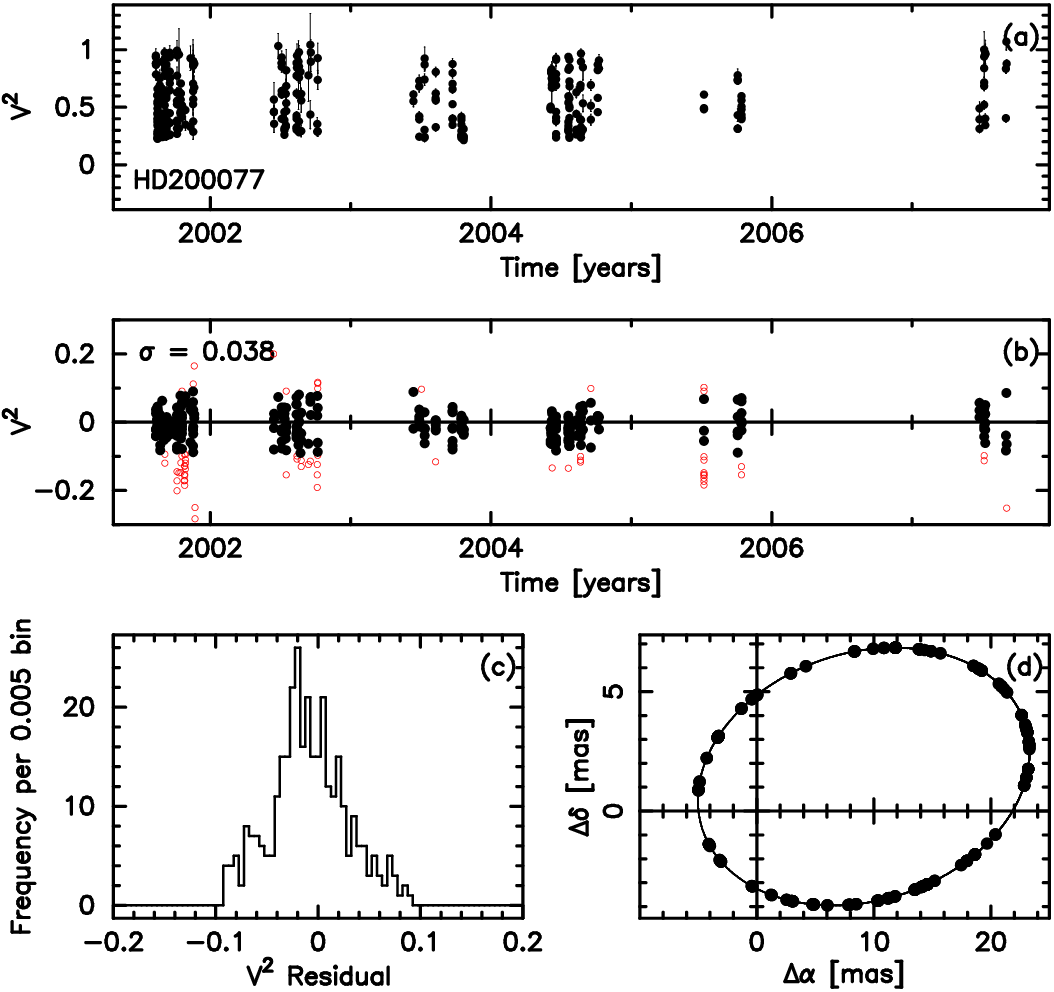}
\caption{Visibility measurements of HD200077 as a function of time (a),
their best-fit residuals as a function of time (b) and histogram (c).
The measurements used to determine the best-fit orbital solution
are denoted with filled circles. The corresponding orbital coverage 
and the relative orbit is shown in the panel (d).}
\end{figure}

%
%

\begin{figure}
\figurenum{14}
\epsscale{0.8}
\plotone{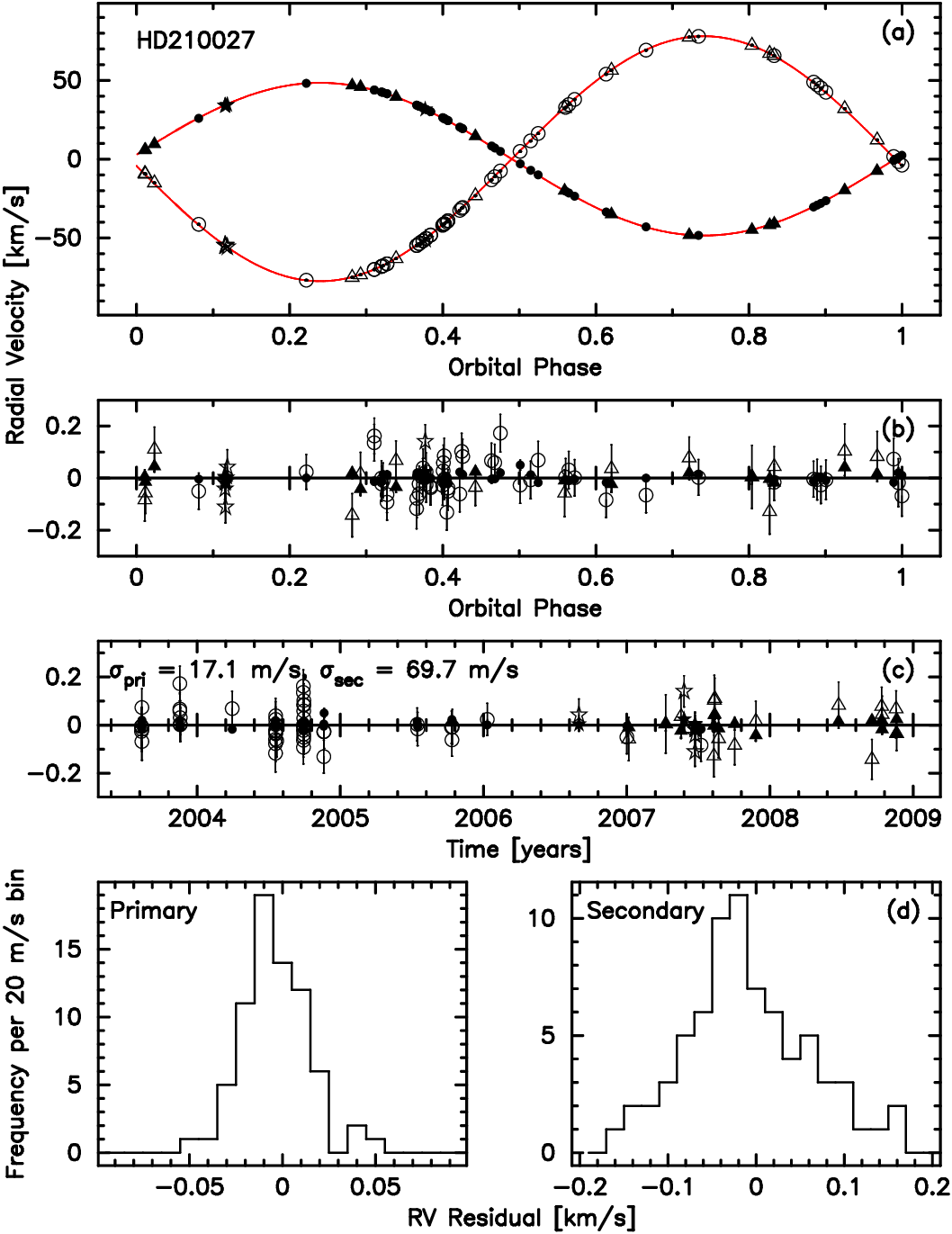}
\caption{Observed and modelled radial velocities of HD210027 as a
function of the orbital phase (a), their best-fit residuals as
a function of the orbital phase (b) and time (c). The histograms
of the residuals for the primary and secondary (d). The Keck I/Hires is
denoted with circles, Shane/CAT/Hamspec with triangles and TNG/Sarg with stars.}
\end{figure}

%
%

\begin{figure}
\figurenum{15}
\epsscale{0.8}
\plotone{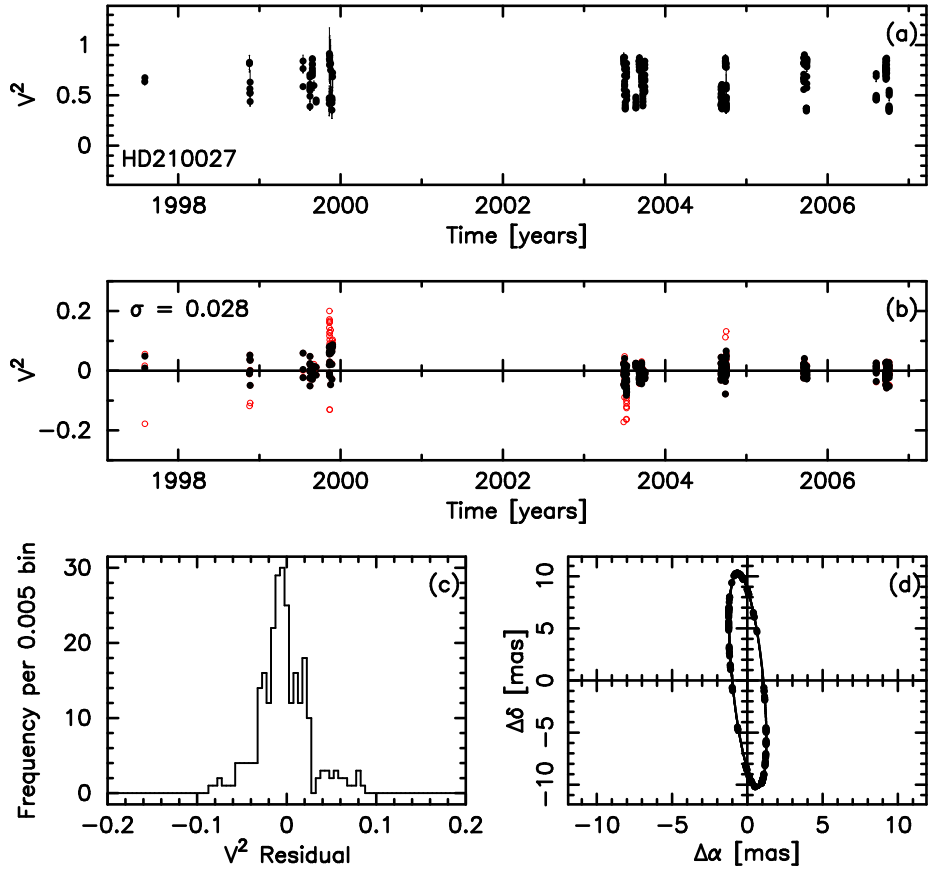}
\caption{Visibility measurements of HD210027 as a function of time (a),
their best-fit residuals as a function of time (b) and histogram (c).
The measurements used to determine the best-fit orbital solution
are denoted with filled circles. The corresponding orbital coverage 
and the relative orbit is shown in the panel (d).}
\end{figure}

%
%

\begin{figure}
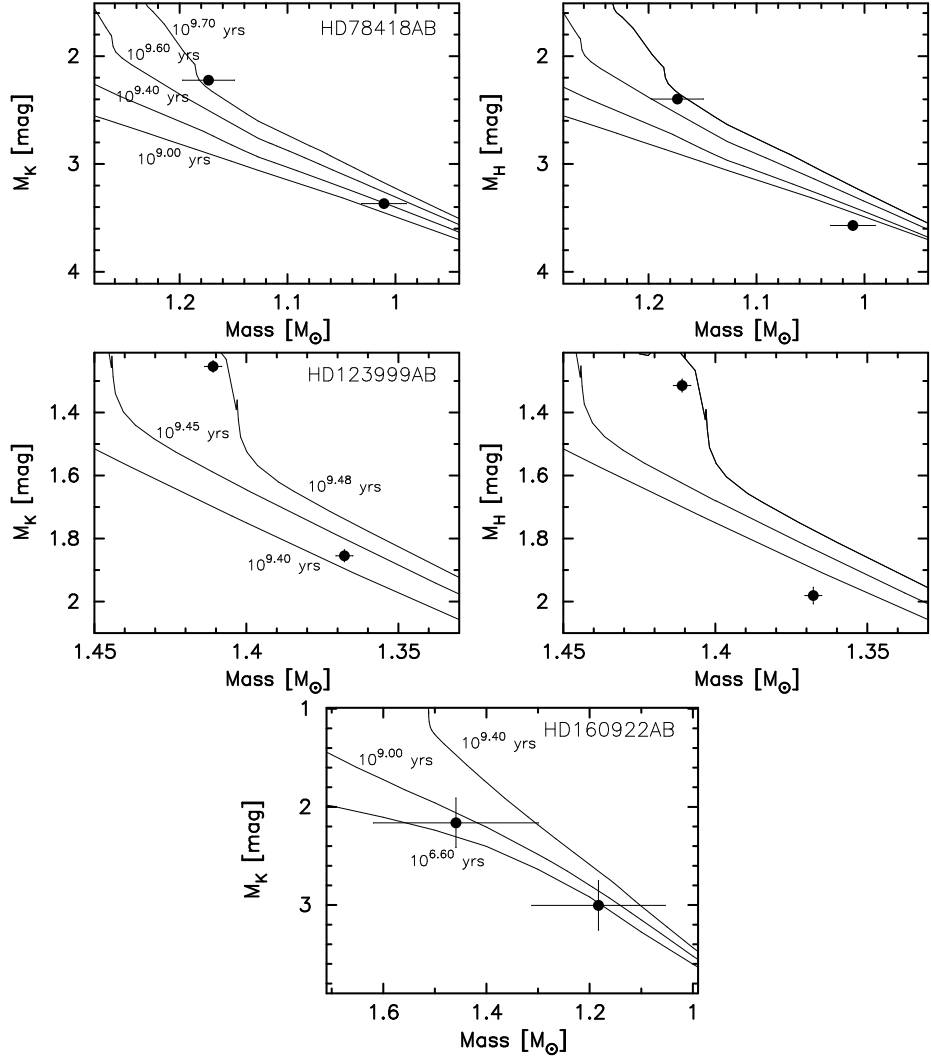

\figurenum{16}
\epsscale{0.8}
\plotone{fig16a.ps}

\plotone{fig16b.ps}

\epsscale{0.4}

\plotone{fig16c.ps}
\caption{The isochrones in the mass - absolute magnitude plane for HD78418
(top), HD123999 (middle) and HD160922 (bottom) together with the
corresponding best-fitting isochrones.}
\end{figure}

%
%

\begin{figure}
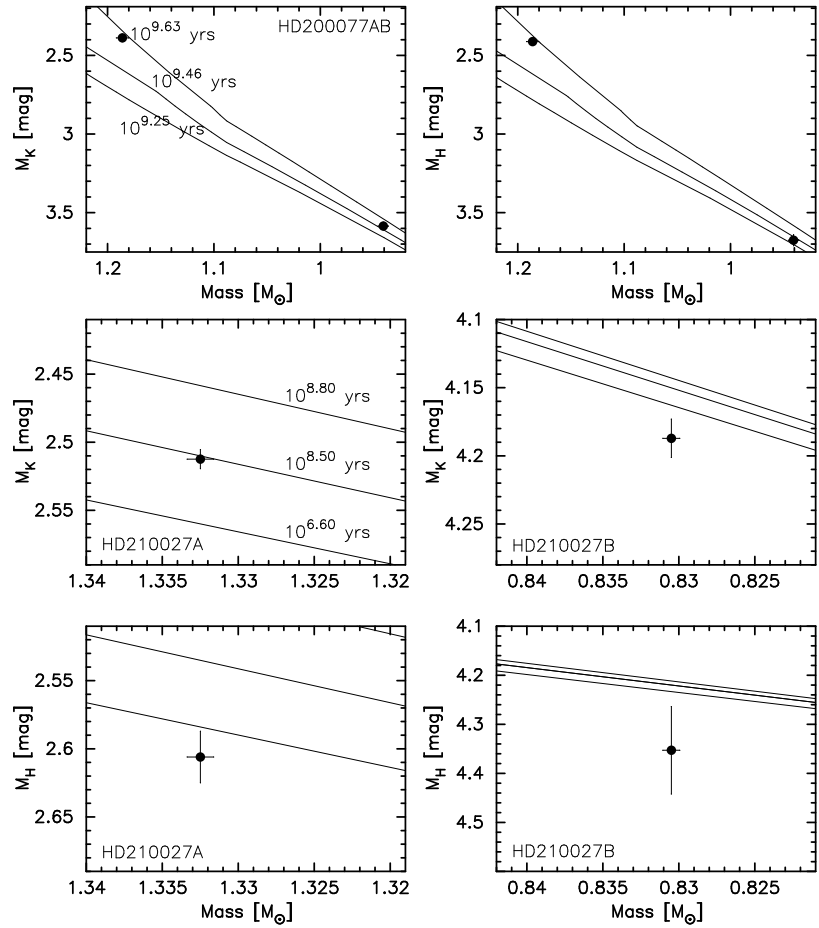

\figurenum{17}
\epsscale{0.7}
\plotone{fig17a.ps}

\plotone{fig17b.ps}
\caption{The isochrones in the mass - absolute magnitude plane for HD200077
(top two panels) and HD210027 (bottom four panels) together with the 
corresponding best-fitting isochrones. In the case of HD210027 their
companions are shown separately so as it is easier to compare the
errorbars with the isochrones.}
\end{figure}

\clearpage

%
%

\begin{figure}
\epsscale{0.85}
\figurenum{18}
\plottwo{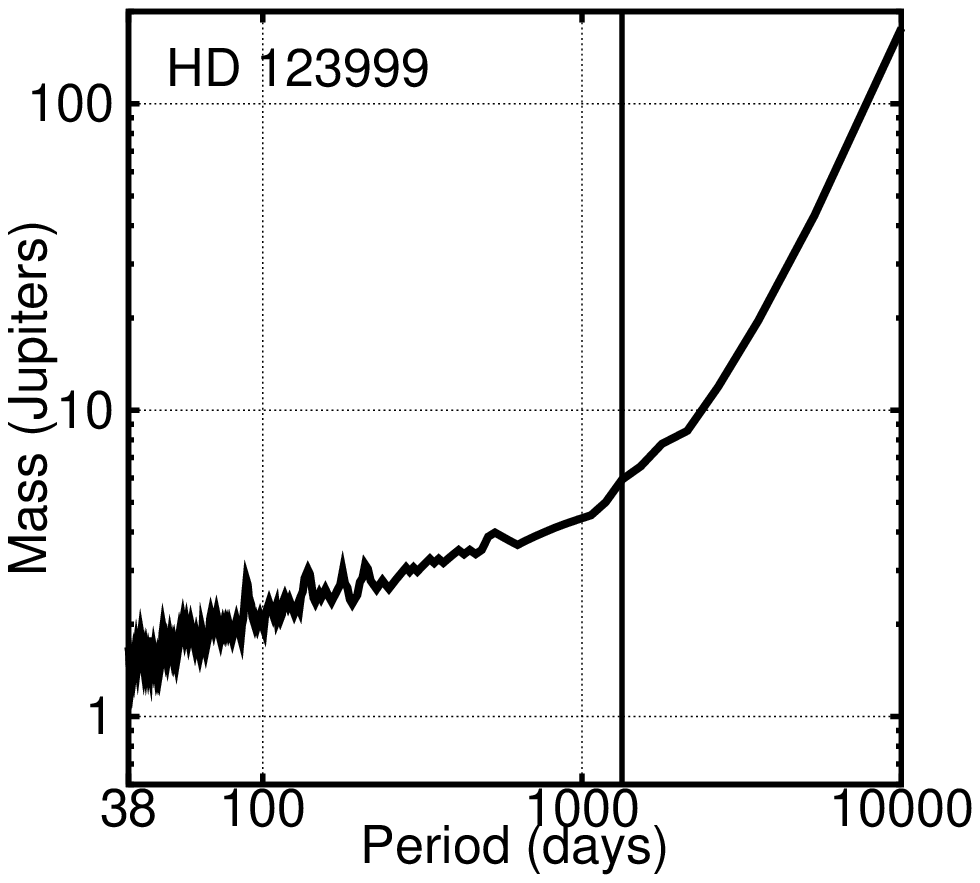}{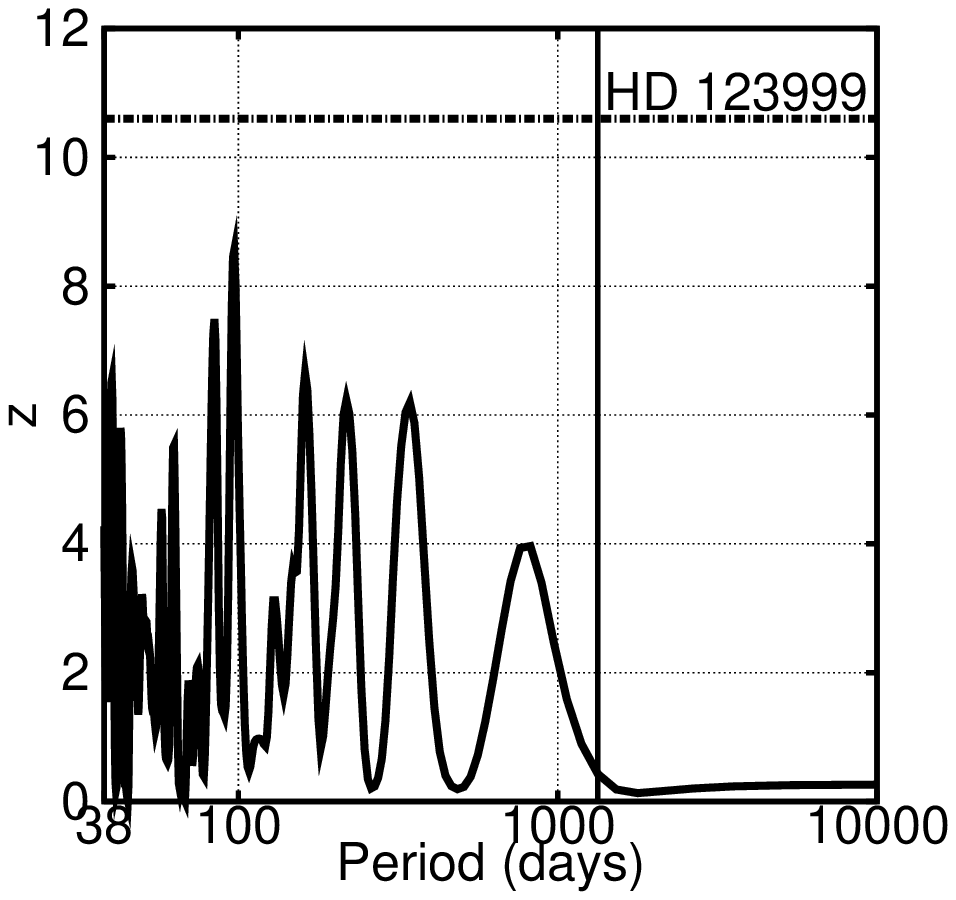}

\plottwo{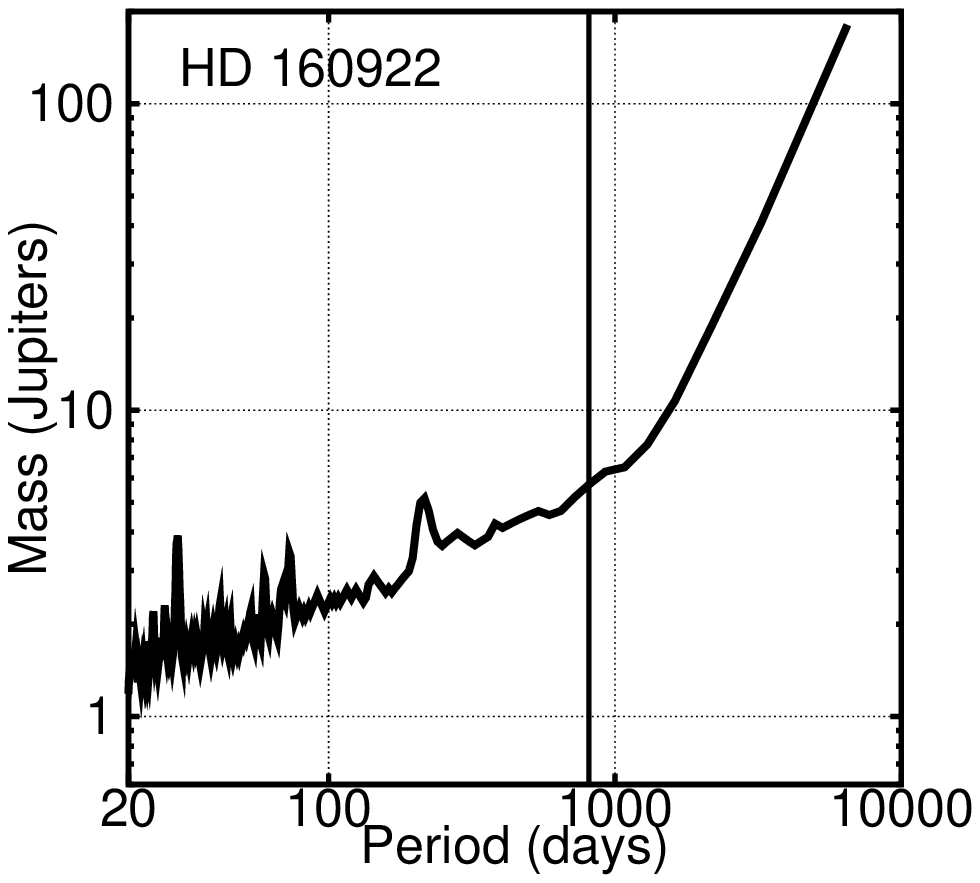}{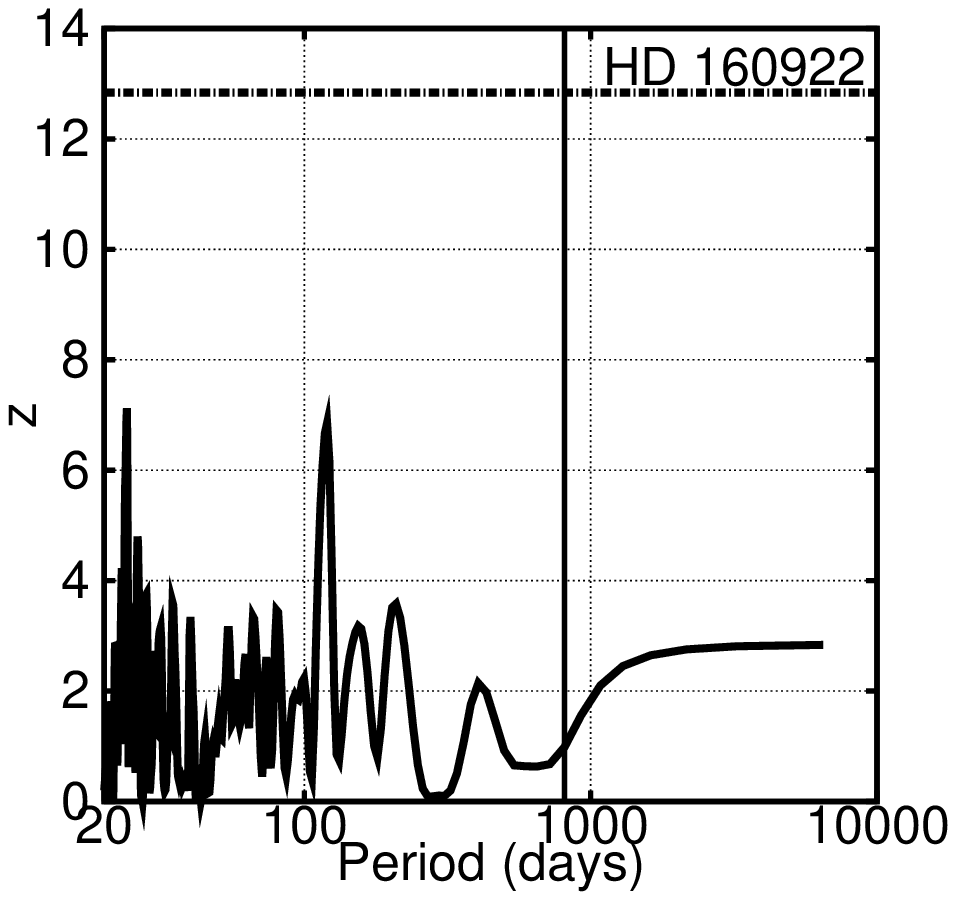}

\plottwo{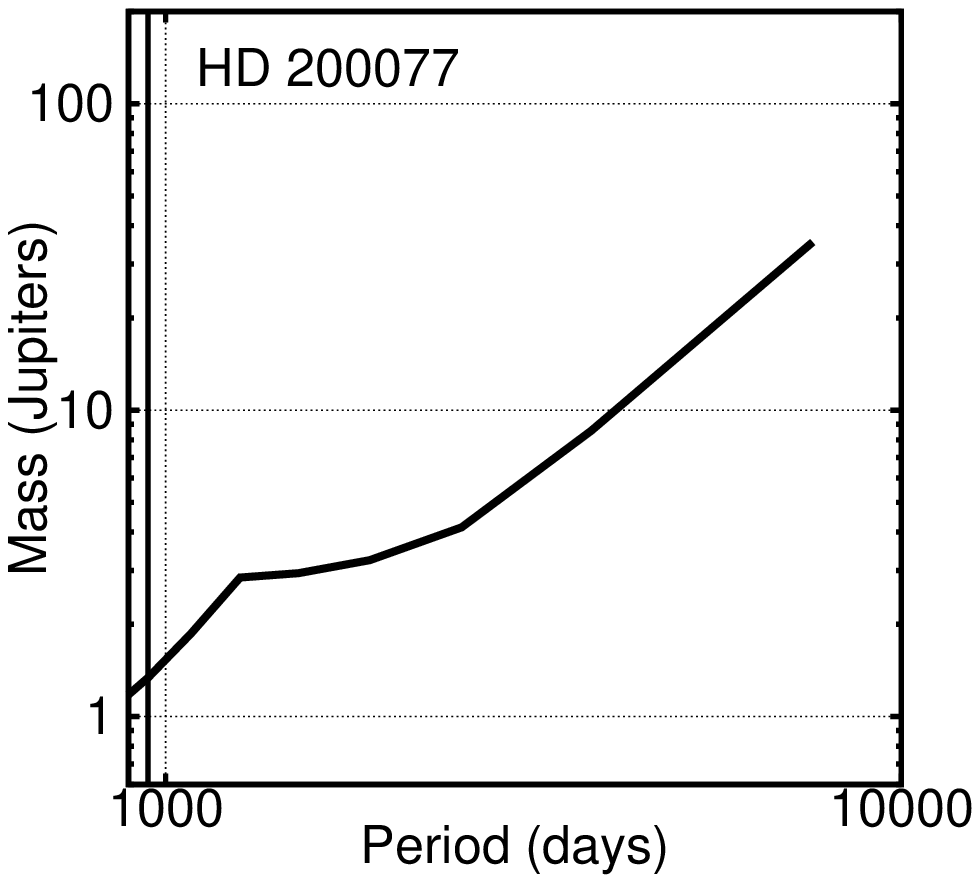}{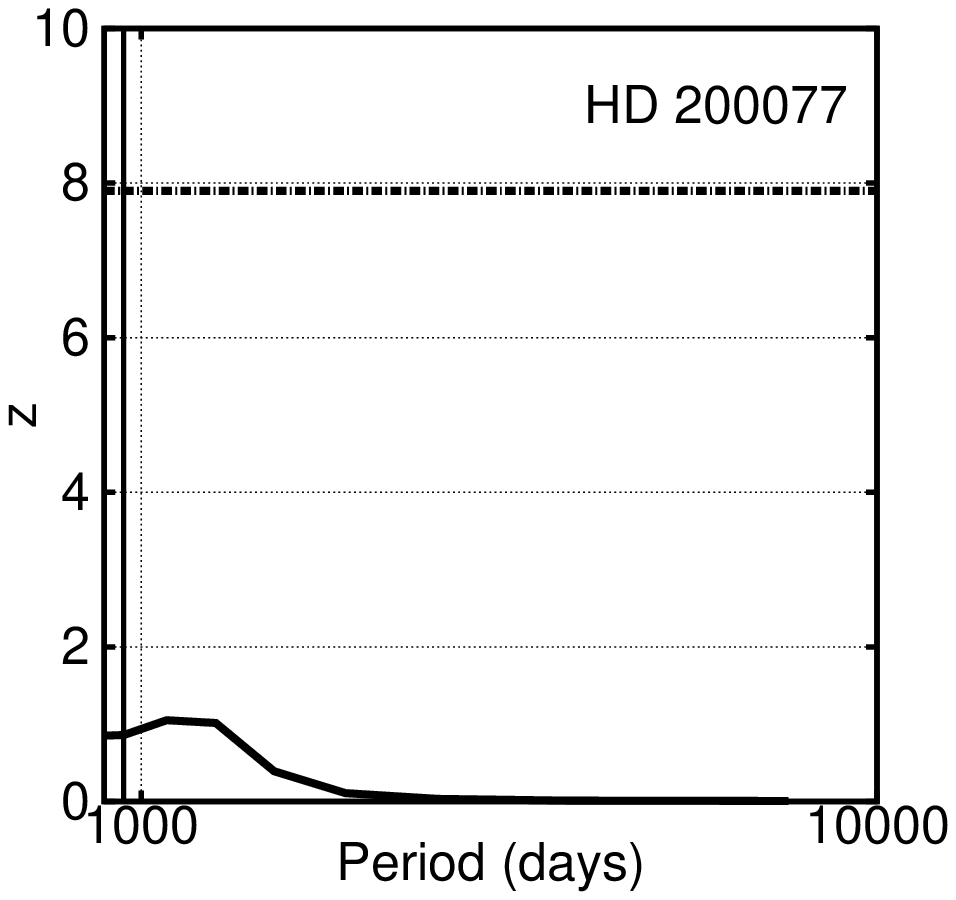}
\caption{Circumbinary planet detection limits in the log P - log M space (orbital 
period -mass; left panels) and periodograms (right panels). The solid
horizontal line in the right panels is a planet detection limit corresponding 
to the 99\% confidence 
level. The vertical line near the orbital period of 1000 days denotes the
time span of the data set. The first stable planetary orbits have
the periods of respectively 38, 20 and 890 days as calculated based
on \cite{Holman:99::}}
\end{figure}

%
%

\begin{figure}
\figurenum{19}
\epsscale{0.7}
\plotone{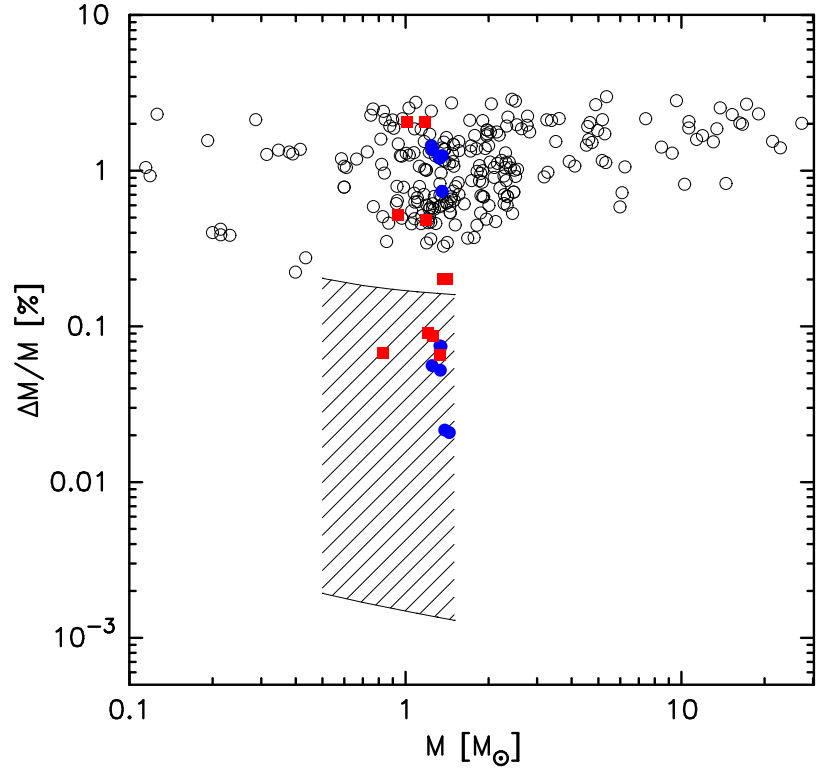}

\caption{Fractional accuracy in masses for the binary stars
with the most accurate mass determinations in the literature. 
Open circles denote the binaries from a recent review 
by \cite{Torres:09::}. Filled (blue) circles denote the double
neutron star systems B1913+16, B2127+11C, B1534+12, J0737-3039,
J1756-2251 and J1906+0746 characterized with radio pulsar timing
\citep{Nice:08::}. Filled (red) rectangles denote the masses 
for HD78418, HD123999, HD200077 and HD210027 from this paper and a mass 
determination for AI Phe from our recent work \citep{Hel:09::}. The hatched 
area is the expected precision in masses for double-lined eclipsing
binary stars assuming a mass range of the components of 0.5-1.5 M$_{\odot}$, 
orbital period range of 3-23 days, orbital inclination range of 
85-90$^{\circ}$, inclination's error range of 0.05-0.3$^{\circ}$
and radial velocity amplitudes' error range of 1-31 m~s$^{-1}$.}
\end{figure}

\clearpage

%
%

\begin{deluxetable}{llrrrrrrrrr}
\tabletypesize{\footnotesize}
\tablecaption{Radial velocities of HD78418, HD123999, HD160922, HD200077 and
HD210027}
\tablewidth{0pt}
\tablehead{ \colhead{Target} & \colhead{Time} & 
\colhead{RV$_1$} & \colhead{$\sigma_1$} & \colhead{O-C$_1$} & \colhead{$\epsilon_1$} 
& \colhead{RV$_2$} & \colhead{$\sigma_2$} & \colhead{O-C$_2$} & \colhead{$\epsilon_2$}
& \colhead{Inst.} \\
\colhead{} & \colhead{(TDB-2400000.5)} & \colhead{(km~s$^{-1}$)} & \colhead{(km~s$^{-1}$)} & 
\colhead{(km~s$^{-1}$)} & \colhead{(km~s$^{-1}$)} & 
\colhead{(km~s$^{-1}$)} & \colhead{(km~s$^{-1}$)} & \colhead{(km~s$^{-1}$)} }
\startdata
HD78418 & 53094.347119 & -14.83066  &  0.00854  &  0.00538 &  0.00299 &
               38.51581  &  0.01243  &  0.00234  & 0.00580 & HO \\
 & 53094.395761 & -14.91605  &  0.01039  &  0.00241 &  0.00663 &
               38.60908  &  0.01267 &  -0.00023 &  0.00628 & HO \\
 & ... & & & & & & &  \\
 & 54101.570433 &  -7.42109  &  0.00876 &  -0.00731 &  0.00357 & 
               29.91678  &  0.01302  &  0.01759 &  0.00697 & HN \\
 & 53744.479316 &  34.57326  &  0.00817  &  0.00352 &  0.00167 &
              -18.81645  &  0.01216  &  0.01278 &  0.00518 & HN \\
 & ... & & & & & & &  \\
 & 54788.464774 &  18.30307  &  0.00696 &  -0.00953 &  0.00485 &
                0.16661 &   0.03029  &  0.01501 &  0.01374 & H \\
 & 54430.494032 &  -2.05150  &  0.00974 &  -0.00216 &  0.00836 &
               23.77923  &  0.03049  &  0.01071 &  0.01416 & H \\
 & ... & & & & & & &  \\
 & 54190.964445  & 34.26932  &  0.01290  &  0.00397 &  0.01290 &
              -19.07279 &   0.01534  &  0.03793 &  0.01534 & S \\
 & 54191.956955 &  29.78440  &  0.01249  & -0.01327 &  0.01249 &
              -13.94505  &  0.01050  & -0.01623 &  0.01050 & S \\
\hline
HD123999 & 53454.573974 & -51.41296  &  0.03155  &  0.00976  &  0.00585  &
 72.66300  &  0.03675 &  0.00268  &  0.01119 & HN \\
 & 53456.545316 &  39.94170  &  0.03231 &  -0.02204  & 0.00911   &
 -21.65504  &  0.03623  & -0.04321 &   0.00937 & HN \\
 & ... & & & & & & &  \\
 & 53978.858598 &  24.88006 &   0.04693  &  0.00628 &  0.03523 &
 -6.43652 &   0.03679 &  -0.01637 & 0.03679 & S \\
 & 54247.021370 &  48.74097 &   0.03697  &  0.07561 & 0.02015 &
-31.00316 &   0.03535 &  -0.04846 & 0.03535 & S \\
 & ... & & & & & & &  \\
 & 54107.560803 & -52.70193  &  0.02118 &  -0.02634 & 0.01388 &
73.95953  &  0.03339 &  -0.02783 &  0.01819 & H \\
 & 54108.572117 & -25.53836 &   0.04058 &  -0.01830 & 0.03730 &
45.98819  &  0.05134 &   0.02694 &  0.04303 & H 
\enddata
\tablecomments{$\sigma_{1,2}$ denote the total errors used in the
least-squares fits and $\epsilon_{1,2}$ denote the internal errors.
The last column denotes the spectrograph used to obtain a
measurement. HN stands for Keck I/Hires with the upgraded detector, HO with
the old detector, S stands for TNG/Sarg and H for Shane/CAT/Hamspec.
The rest of this data set and the other RV sets are available via the link 
to the machine-readable version.}
\end{deluxetable}


%
%

\begin{deluxetable}{lccccc}
\footnotesize
\tablewidth{0pt}
\tablecaption{Assumed Parameters for HD78418, HD123999, HD160922,
HD200077 and HD210027\tablenotemark{a}}
\tablehead{\colhead{Parameter} & \colhead{HD78418} & \colhead{HD123999} &
\colhead{HD160922} & \colhead{HD200077} & \colhead{HD210027} }
\startdata
Eff. temperature, primary, $T_1$ (K)\dotfill           & 6000  & 6130  & 
6500 & 6000 & 6642 \\ 
Eff. temperature, secondary, $T_2$ (K)\dotfill         & 5900  & 6230  &
5900 & 5500 & 4991 \\ 
Potential, primary, $\Omega_1$\dotfill                 & 27.1  & 12.03 &
14.3 & 26.7 & 20.0 \\ 
Potential, secondary, $\Omega_2$\dotfill               & 34.9  & 15.21 &
14.1 & 26.5 & 22.5 \\ 
Synchronization factor, primary, $F_1$\dotfill         & 1.51  & 1.5  &
1.0 & 6.57 & 1.0 \\
Synchronization factor, secondary, $F_2$\dotfill       & 1.51  & 1.5  &
1.0 & 6.57 & 1.0 \\
Gravity darkening exponent, primary, $g_1$\dotfill     & 0.4   & 0.35  & 
0.3 & 0.35 & 0.3 \\
Gravity darkening exponent, secondary, $g_2$\dotfill   & 0.4   & 0.35  & 
0.3 & 0.4 & 0.4 \\
Albedo, primary, $A_1$\dotfill                         & 0.5   & 0.5   & 
0.5 & 0.5 & 0.5 \\
Albedo, secondary, $A_2$\dotfill                       & 0.5   & 0.5   &
0.5 & 0.5 & 0.5 \\
Metallicity \dotfill                                   & -0.09 & 0.0   &
0.0  & -0.14 & 0.0 \\
Apparent diameter, primary, $\theta_1$ (mas)\dotfill   & 0.45  & 0.638 &
0.5 & 0.26 & 1.06 \\
Apparent diameter, secondary, $\theta_2$ (mas)\dotfill & 0.30  & 0.480 &
0.4 & 0.21 & 0.6 
\enddata
\tablenotetext{a}{The more accurate parameters for HD123999 and HD210027
come from respectively \cite{Bod:05::} and \cite{Mo:00::}. The remaining ones
are our best estimates based on the available information about the 
targets.}
\end{deluxetable}


\clearpage

%
%

\begin{landscape}

%
%

\begin{deluxetable}{llcccc}
\footnotesize
\tablewidth{0pt}
\tablecaption{Calibration stars for visibility measurements of
HD78418, HD123999, HD160922, HD200077 and HD210027\tablenotemark{a} }
\tablehead{\colhead{Target} & \colhead{Calibrator} & \colhead{Spectral} &  
\colhead{Magnitude} & \colhead{Angular Separation} & \colhead{Apparent} \\  
\colhead{} & \colhead{} & \colhead{Type} &  \colhead{}
& \colhead{from Target (deg)} & \colhead{Diameter (mas)} }
\startdata
HD78418 & & & & & \\  
 & HD79452 & G6III & 6.0 V, 3.8 K & 8.1 & 0.79 \\
 & HD73192 & K2III & 6.0 V, 3.3 K & 9.0 & 1.38 \\
 \hline
HD123999 & & & & & \\
 & HD121107 & G5III & 5.7 V, 3.6 K & 8.2 & 0.70 \\
 & HD128167 & F3V   & 4.5 V, 3.6 K & 7.1 & 0.79 \\
 & HD123612 & K5III & 6.6 V, 3.0 K & 0.9 & 1.29 \\
 \hline
HD160922 & & & & & \\
 & HD154633 & G5V & 6.1 V, 4.5 K & 5.4 & 0.31 \\
 & HD158633 & K0V & 6.4 V, 4.4 K & 1.8 & 0.59 \\
 & HD168151 & F5V & 5.0 V, 3.9 K & 5.7 & 0.56 \\
 \hline
HD200077 & & & & & \\
 & HD192640 & A2V   & 4.9 V, 4.9 K & 9.5 & 0.71 \\
 & HD192985 & F5V   & 5.9 V, 4.8 K & 9.6 & 0.39 \\
 & HD199763 & G9III & 6.5 V, 4.3 K & 9.9 & 0.78 \\
 \hline
HD210027 & & & & & \\
 & HD211006 & K2III & 5.9 V, 3.2 K &  3.6 & 1.06 \\
 & HD211432 & G9III & 6.4 V, 4.2 K &  3.2 & 0.70 \\
 & HD215510 & G6III & 6.3 V, 4.2 K & 10.7 & 0.85 \\
 & HD210459 & F5III & 4.3 V, 2.5 K &  7.9 & 0.81 
\enddata
\tablenotetext{a}{The adopted diameters of the calibrators are determined
from their effective temperature and bolometric flux derived from archival
photometry using {\tt getCal} tool (ver. 2.10.7) supplied by the NASA Exoplanet
Science Institute.}
\end{deluxetable}

\begin{deluxetable}{lccccc}
\tablewidth{0pt}
\tabletypesize{\footnotesize}
\tablecaption{Best-fit Orbital Solutions for HD78418, HD123999, HD160922,
HD200077 and HD210027\tablenotemark{a}}
\tablehead{\colhead{Parameter} & \colhead{HD78418} & \colhead{HD123999} &
\colhead{HD160922} & \colhead{HD200077} & \colhead{HD210027} }
\startdata
Apparent semi-major axis, $\hat{a}$ (mas) \dotfill          & 5.8696(96)    & 
3.4706(55) & 3.469(17)         & 14.453(18)  & 10.329(16)  \\
Period, $P$ (d) \dotfill                                    & 19.412347(23) & 
9.6045601(36) & 5.2797766(44)     & 112.5131(14)  &  10.2130253(16) \\
Time of periastron, $T_p$ (TDB-2400000.5) \dotfill                    & 53895.4025(24)&
54099.93572(70) & 54348.583(83)     & 53830.168(14)  & 52997.378(52)  \\
Eccentricity, $e$ \dotfill                                  & 0.19494(11)   &
0.19214(15) & 0.00220(31)       & 0.66227(51)  & 0.001764(63)  \\
Longitude of the periastron, $\omega$ (deg) \dotfill        & 283.389(39)   &
286.832(29) & 314.8(5.6)        & 197.071(25)  & 272.8(1.8)  \\
Longitude of the ascending node, $\Omega$ (deg) \dotfill    & 171.892(85)   &
80.49(10) & 1.23(32)        & 89.403(28)  &   176.262(75) \\
Inclination, $i$ (deg) \dotfill                             & 146.88(25)    &
107.95(12) & 151.4(1.1)        & 118.681(80)  &  95.83(12) \\
Magnitude difference (K band), $\Delta K$ \dotfill          & 1.1445(131)   &
0.601(13) &  0.841(18)  & 1.1968(88)  & 1.675(15)  \\
Magnitude difference (H band), $\Delta H$ \dotfill          & 1.1726(349)   &
0.667(30) & ...  & 1.263(38)  &  1.75(11) \\
Velocity amplitude of the primary, $K_1$ (km~s$^{-1}$) \dotfill    & 26.4961(35)   &
67.189(11) & 36.254(16)        & 29.373(82)  &  48.4757(39) \\
Velocity amplitude of the secondary, $K_2$ (km~s$^{-1}$) \dotfill  & 30.7579(65)   &
69.311(14) & 44.720(16)        & 37.03(11)  &  77.777(16) \\
Gamma velocity, $V_{R}$ (km~s$^{-1}$) \dotfill  & 9.7478(60) & 9.646(13) &
-14.011(19) & -36.009(18) &-4.6504(34)
\enddata
\tablenotetext{a}{The numbers in parentheses are the 1-sigma errors in the last
digits quoted.}
\end{deluxetable}

\clearpage

%
%

\begin{deluxetable}{lccccc}
\footnotesize
\tablewidth{0pt}
\tabletypesize{\scriptsize}
\tablecaption{Various Fit Parameters for HD78418, HD123999, HD160922,
HD200077 and HD210027\tablenotemark{a}}
\tablehead{\colhead{Parameter} & \colhead{HD78418} & \colhead{HD123999} &
\colhead{HD160922} & \colhead{HD200077} & \colhead{HD210027} }
\startdata
{\it Velocity offsets} & & & & & \\
Secondary vs primary (km~s$^{-1}$) \dotfill       & 0.239(12) & 0.032(21) & 
0.116(29) & 0.340(42) & 0.333(16) \\
Hires old vs new detector, primary, $v_{11}$ (km~s$^{-1}$) \dotfill       &
-0.0009(71) & ... & ... & ... & -0.0053(38)\\
Hires old vs new detector, secondary, $v_{12}$ (km~s$^{-1}$) \dotfill     &
-0.0131(11) & ... & ... & ... & 0.003(21)\\
Hamspec vs Hires new detector, primary, $v_{21}$ (km~s$^{-1}$) \dotfill   & 
0.0358(67)  & -0.010(15) & ... & ... & -0.0141(68)\\
Hamspec vs Hires new detector, secondary, $v_{22}$ (km~s$^{-1}$) \dotfill & 
0.054(12)   & 0.042(18)  & ... & ... & 0.026(27)\\
Sarg vs Hires new detector, primary, $v_{31}$ (km~s$^{-1}$) \dotfill      & 
-0.3059(87) & -0.182(24) & ... & -0.3092(79) & -0.3026(93)\\
Sarg vs Hires new detector, secondary, $v_{32}$ (km~s$^{-1}$) \dotfill    & 
-0.286(10)  & -0.151(25) & ... & -0.229(23) & -0.256(32)\\
Hamspec vs Sarg, primary, $v_{41}$ (km~s$^{-1}$) \dotfill   &
...  & ... & 0.262(21) & ... & ... \\
Hamspec vs Sarg, secondary, $v_{42}$ (km~s$^{-1}$) \dotfill &
...   & ... & 0.344(21) & ... & ... \\
 & & & & & \\
{\it Least-squares fit parameters} & & & & & \\
Number of RV measurements, total \dotfill             & 58 & 64 & 36  & 26 & 144\\
Number of RV measurements, Keck/Hires \dotfill        & 26 & 16 & ... & 18 & 102 \\
Number of RV measurements, Shane/CAT/Hamspec \dotfill & 24 & 34 & 20  & ...& 30\\
Number of RV measurements, TNG/Sarg \dotfill          &  8 & 14 & 16  & 8  & 12\\
Number of V$^2$ measurements \dotfill & 415 & 346 & 61 & 329 & 266\\

Additional RV error, Keck/Hires, primary/secondary (m~s$^{-1}$) \dotfill & 
8.0/11.0 & 31.0/35.0 & ... & 5.0/26.0 & 10.5/65.0\\
Additional RV error, Shane/CAT/Hamspec, primary/secondary (m~s$^{-1}$) \dotfill &
5.0/27.0 & 16.0/28.0 & 27.0/15.0 & ... & 18.5/68.0\\
Additional RV error, TNG/Sarg, primary/secondary (m~s$^{-1}$) \dotfill   & 
0.0/0.0  & 31.0/0.0 & 37.0/25.0 & 0.0/26.0 & 10.5/38.0\\

Combined RV rms, primary/secondary (m~s$^{-1}$) \dotfill & 13.1/24.3 &
34.0/38.3 & 37.9/31.9 & 8.5/30.4 & 17.1/69.7\\
Keck/Hires RV rms, primary/secondary (m~s$^{-1}$) \dotfill & 6.9/14.3 &
30.2/35.8 & ... & 6.7/17.1 & 14.1/64.6 \\
Shane/CAT/Hamspec RV rms, primary/secondary (m~s$^{-1}$) \dotfill & 18.2/33.8 & 
36.3/37.6 & 25.6/14.8 & ... & 22.7/84.2 \\
TNG/Sarg RV rms, primary/secondary (m~s$^{-1}$) \dotfill   & 8.2/17.4 &
40.2/26.8 & 35.9/37.1 & 10.4/25.9 & 12.3/55.8 \\

V$^2$ rms \dotfill & 0.0351 & 0.0261 & 0.0274 & 0.0381 & 0.0281\\
RV $\chi^2$, primary/secondary \dotfill & 31.60/32.99 & 28.21/28.20 &
19.31/18.19 & 11.28/13.41 & 70.40/64.94\\
V$^2$ $\chi^2$ \dotfill & 433.71 & 482.46 & 50.59 & 522.01 & 248.29\\
Degrees of freedom, $DOF$ \dotfill & 454 & 393 & 83 & 355 & 391\\
Total reduced $\chi^2$, $\chi^2/DOF$ \dotfill                     & 1.096         &
1.371 & 1.061 &  1.608  &  0.981 
\enddata
\tablenotetext{a}{The numbers in parentheses are the 1-sigma errors in the last
digits quoted.}
\end{deluxetable}
\clearpage
\end{landscape}

\clearpage

%
%

\begin{deluxetable}{llcc}
\tablewidth{0pt}
\tablecaption{Physical Parameters for HD78418, HD123999, HD160922, HD200077 and
HD210027\tablenotemark{a}}
\tablehead{\colhead{Star} & \colhead{Parameter} & \colhead{Primary} &  \colhead{Secondary}}
\startdata
HD78418 & Semi-major axis, $a_{1,2}$ (AU)\dotfill  & 0.084874(11) & 0.098526(21)\\
 & M $\sin^3i$ (M$_{\odot}$)\dotfill & 0.191350(50) & 0.164836(45) \\ 
 & Mass, $M$ (M$_{\odot}$)\dotfill          & 1.173(24)    & 1.011(21)\\
 & M$_{K,2MASS}$ (mag)\dotfill              & 2.223(16)    & 3.367(18) \\
 & M$_{H,2MASS}$ (mag)\dotfill              & 2.397(37)    & 3.570(45)\\
 & Parallax, $\kappa$ (mas)\dotfill         &  & \hspace{-2.5cm}{32.004(52)} \\
 & Distance, $d$ (pc)\dotfill               &  & \hspace{-2.5cm}{31.246(51)} \\
HD123999 & Semi-major axis, $a_{1,2}$ (AU)\dotfill  & 0.061193(10) & 0.063125(13)\\
 & M $\sin^3i$ (M$_{\odot}$)\dotfill        & 1.21467(33) & 1.17749(33) \\ 
 & Mass, $M$ (M$_{\odot}$)\dotfill          & 1.4109(28)  & 1.3677(28) \\
 & M$_{K,2MASS}$ (mag)\dotfill              & 1.254(18)   & 1.855(19)\\
 & M$_{H,2MASS}$ (mag)\dotfill              & 1.314(20)   & 1.981(26)\\
 & Parallax, $\kappa$ (mas)\dotfill         &  & \hspace{-2.5cm}{27.917(44)}\\
 & Distance, $d$ (pc)\dotfill               &  & \hspace{-2.5cm}{35.820(57)}\\
HD160922 & Semi-major axis, $a_{1,2}$ (AU)\dotfill  & 0.036728(16) & 0.045306(17)\\
 & M $\sin^3i$ (M$_{\odot}$)\dotfill        & 0.160409(88) & 0.130039(80) \\ 
 & Mass, $M$ (M$_{\odot}$)\dotfill          & 1.46(16)     & 1.18(13) \\
 & M$_{K,2MASS}$ (mag)\dotfill              & 2.16(25)     & 3.00(25) \\
 & M$_{H,2MASS}$ (mag)\dotfill              &  ... & ...\\
 & Parallax, $\kappa$ (mas)\dotfill         &  & \hspace{-2.5cm}{42.29(27)}\\
 & Distance, $d$ (pc)\dotfill               &  & \hspace{-2.5cm}{23.65(15)}\\
HD200077 & Semi-major axis, $a_{1,2}$ (AU)\dotfill  & 0.25945(74) & 0.32711(96)\\
 & M $\sin^3i$ (M$_{\odot}$)\dotfill        & 0.8008(34) & 0.6352(30) \\ 
 & Mass, $M$ (M$_{\odot}$)\dotfill          & 1.1860(57)  & 0.9407(49)\\
 & M$_{K,2MASS}$ (mag)\dotfill              & 2.389(24)   & 3.585(25)\\
 & M$_{H,2MASS}$ (mag)\dotfill              & 2.412(23)   & 3.675(36)\\
 & Parallax, $\kappa$ (mas)\dotfill         &  & \hspace{-2.5cm}{24.640(32)}\\
 & Distance, $d$ (pc)\dotfill               &  & \hspace{-2.5cm}{40.571(52)}\\
HD210027 & Semi-major axis, $a_{1,2}$ (AU)\dotfill  & 0.0457446(37) & 0.073395(15)\\
 & M $\sin^3i$ (M$_{\odot}$)\dotfill        & 1.31190(29) & 0.81767(22)\\ 
 & Mass, $M$ (M$_{\odot}$)\dotfill          & 1.33249(86) & 0.83050(55)\\
 & M$_{K,2MASS}$ (mag)\dotfill              & 2.5125(70) & 4.187(14)\\
 & M$_{H,2MASS}$ (mag)\dotfill              & 2.606(19) & 4.353(89)\\
 & Parallax, $\kappa$ (mas)\dotfill         &  & \hspace{-2.5cm}{86.70(14)} \\
 & Distance, $d$ (pc)\dotfill               &  & \hspace{-2.5cm}{11.534(19)} 
\enddata
\tablenotetext{a}{The numbers in parentheses are the 1-sigma errors in the last
digits quoted.}
\end{deluxetable}

\end{document}